\documentclass[a4paper,12pt,twoside,titlepage]{report}

\usepackage{amssymb,amsthm}
\usepackage{euscript} 
\usepackage{txfonts}
\usepackage[dvips]{graphicx}
\usepackage[all]{xy}
\usepackage{fancyhdr}

\usepackage{style}

\title{Some comments on the local curve\\(B-side)}
\author{Antonio Ricco\\ {\small International School for Advanced Studies %
                                (SISSA/ISAS), Trieste, Italy}}
\date{}

\begin{document}

\thispagestyle{empty}

\begin{center}
\begin{tabular}{c}
\\
\\
\\
\\
\\
{\LARGE \textbf{Some comments on the local curve}} \\
{\small}\\
{\LARGE \textbf{(B-side)} }\\
\\
{\large Antonio Ricco}\\
\\
\\
\\
\\
\\
\\
\\
\begin{tabular}{ll}
{\large Supervisors:$\qquad\qquad\qquad\qquad$}& {\large Referee:}\\
&\\
 {\large Prof.~Loriano Bonora}& {\large Prof. Marcos Mari\~no}\\  
  {\large Prof.~Ugo Bruzzo}   & \\
\end{tabular}
\end{tabular}

\begin{tabular}{c}
\\
\\
\\
\\
\\
\\
\\
\\
\\
\\
\\
Dissertation submitted in partial fulfillment of the\\
requirements for the degree of Doctor of Philosophy\\
\\
Mathematical Physics Sector\\
International School for Advanced Studies\\
Trieste\\
\\
2006\\
\end{tabular}

\end{center}
\newpage

\begin{tabular}{l}
Antonio Ricco \\
Mailbox 1511 \\
Center of Mathematical Sciences \\
Zhejiang University \\
Hangzhou 310027 \\
E-mail: {\tt ricco@sissa.it}

\end{tabular}

$\ $ \thispagestyle{empty}
\newpage

\thispagestyle{empty}
\begin{center}
$\ $\\
$\ $\\
$\ $\\

{\sl To Her, still a mystery to me.}
\end{center}

$\ $ \thispagestyle{empty}
\newpage
\pagestyle{plain}
\thispagestyle{empty}
$\ $   
\newpage  
\pagenumbering{scroman}
\setcounter{page}{5}

\thispagestyle{empty}
\vspace{8cm}
Every human product, as small as it may be, is of course a social and collective product. 
Having said that, it is a real pleasure for me to express my personal gratitude to all those persons that directly made this thesis possible with their explanations and suggestions.

I thank Loriano Bonora for introducing me to the theory of strings and of topological strings and for inspiring 
and guiding the present work; Ugo Bruzzo for his invaluable help in 
forming my mathematical background, and more generally for his explanations about science, research and life; Giulio Bonelli for his great collaboration and for his continuos suggestions and encouragements;
SISSA and the Mathematical Physics Sector for having been an exciting scientific environment.

I acknowledge stimulating and inspiring discussions with: 
Mina Aganagic,
Elena Andreini,
Carolina Araujo,
Nathan Berkovits,
Stefanella Boatto,
Henrique Bursztyn,
Michele Cirafici,
Robbert Dijkgraaf,
Yassir Dinar,
Boris Dubrovin,
Eduardo Esteves,
Jarah Evslin,
Gregorio Falqui,
Barbara Fantechi,
Bo Feng,
Antonio Grassi,
Tamara Grava,
Luca Griguolo,
Christopher Hull,
Anton Kapustin,
Igor Klebanov,
Kenichi Konishi,
Giovanni Landi,
Kefeng Liu,
Carlo Maccaferri, 
Emanuele Macr\`\i,
Etienne Mann,
Marcos Mari\~{n}o,
Luca Martucci,
Luca Mazzucato,
Alessandro Michelangeli,
Kumar Narain,
Nikita Nekrasov,
Francesco Noseda,
Marco Pacini,
Tony Pantev,
Sara Pasquetti,
Nico Pitrelli,
Cesare Reina,
Moritz Reuter,
Ricardo Schiappa,
Alessio Serafini,
Eric Sharpe,
Leonardo Spanu,
Alessandro Tanzini,
Richard Thomas,
George Thompson,
Alessandro Tomasiello,
Ramadas Trivandrum,
Walter van Suijlekom,
Marcel Vonk,
Jorge Zubelli. 
They were really very important for me in building my scientific understanding. 
I apologize for the many others that I am sure I forgot to mention.

Special thanks go to IMPA, Instituto Nacional de Matem\'atica Pura e Aplicada, and to the ALFA, Am\'erica Latina -- Formaci\'on Acad\'emica, program, for the hospitality I enjoyed in Rio de Janeiro during the period in which this thesis was actually written.

I thank IMS, Institute for Mathematical Sciences, at Imperial College, and IHES, Institut des Hautes \'Etudes  Scientifiques, for their hospitality and for the opportunity to present this work in London and in Bures-sur-Yvette.

\thispagestyle{empty}
\newpage
$\ $ \thispagestyle{empty}
\newpage

\tableofcontents

\chapter*{Introduction}
\addcontentsline{toc}{chapter}{Introduction}





String theory is widely considered to be the leading contender for a reconciliation of general relativity with quantum mechanics. Many propose it as the ``final theory'' or the ``theory of everything''. Even without assuming this viewpoint in its entirety, string theory can certainly provide a deep insight on what one should expect from a theory of quantum gravity, like the calculation of the entropy of a quantum black hole \cite{Strominger:1996sh}, the topological fluctuations of spacetime \cite{Greene:1995hu} and its ``foamy'' nature \cite{Okounkov:2003sp,Iqbal:2003ds}. From a mathematical viewpoint, string theory is continuously giving new intuitions and tools to different areas such as geometry and the theory of dynamical systems.

In spite of these successes, and leaving aside the obvious aim of making string theory face up to reality, its very definition can be considered quite unsatisfactory. Physical quantities, such as partition functions or scattering amplitudes, are defined via a perturbative, asymptotic expansion around a ``classical vacuum,'' the \emph{genus expansion}. This expansion is given in terms of conformal field theories on Riemann surfaces and is not generated in any obvious way by a path-integral of the form used for quantum field theories.

String field theory, both in its open \cite{Witten:1985cc} and in its closed version \cite{Zwiebach:1992ie}, is an attempt to do exactly this, that is, to capture the genus expansion in terms of Feynman graphs.
However, like for quantum field theories, it is clear that many effects are invisible at a perturbative level. 
The understanding that other higher-dimensional objects, called D-branes, belong to the space of states of string theory \cite{Polchinski:1995mt}, and that string theory is ultimately not just a theory of strings, represents an important step in the attempt to give an answer to the question ``what is string theory?,'' i.e. give a nonperturbative definition of string theory. In the perturbative worldsheet formulation D-branes arise as boundary conditions for open strings and, in fact, the ``D'' in the name stands for ``Dirichlet'' boundary conditions for the open strings.
In string field theory, and in its low-energy (super)gravity limit, they can be understood as charged solitons.

In the moduli space of string theory vacua, topological strings have a privileged role. 
The simplification of their degrees of freedom, with respect to e.g. type II strings, permits in many cases to calculate their solution explicitly. Thus, topological strings can be considered a ``theoretical laboratory'' for general features of string theory, as brane dynamics, open/closed string dualities, etc. 
Their connection with algebraic properties of target spaces, in particular for threefolds with trivial canonical bundle (Calabi-Yau spaces), is a source of intuition and inspiration to algebraic geometry. 
Their connection with integrability, both classical and quantum, can, in principle, provide new kinds of integrable systems and integrable hierarchies.
Finally, topological string amplitudes are directly connected to supersymmetry-protected quantities in type II strings and supersymmetric gauge theories. 
These features explain the strong interest in obtaining a deeper understanding of the properties of topological strings. 

In topological strings there are two models, called A-model and B-model. 
As conformal field theories on the worldsheet, 
they are string theories of bosonic type; for the definition of a bosonic string vacuum, see
\cite{Kimura:1993ea,Bershadsky:1993cx} and the lecture notes \cite{Dijkgraaf:1997ip}. 
When defined on Calabi--Yau threefolds, they depend respectively on the K\"ahler structure and on the complex structure of the target space.
For this reason, topological strings could be considered as semi-topological theories of gravity in six dimensions;
for the basics, see the lecture notes \cite{Vonk:2005yv} and the references therein. 
In fact, as suggested by many examples, topological strings could provide a unifying description of different areas of mathematics, such as random objects (matrix models, random partitions), integrable hierarchies, moduli spaces of algebraic objects and enumerative invariants, homological algebra of derived categories.
The connections among these areas stem from a string theory viewpoint as ``dualities,'' namely equivalences of theories defined around different classical vacua. 

One very well-known duality is mirror symmetry, which relates the topological A-model on a Calabi--Yau manifold to the B-model on a dual ``mirror manifold'', mapping the K\"ahler structure of the first to the complex structure of the second. From the target-space viewpoint, it is actually a perturbative duality since it is conjectured to hold order by order in the genus perturbative expansion, but it is a non-perturbative duality seen as a duality of the two $\sigma$-models. Since the predictions about rational curves on the quintic three\-fold \cite{Candelas:1990rm}, mirror symmetry is mathematically much better understood, see for example \cite{Strominger:1996it,Cox:2000vi,Kapustin:2003kt} for different viewpoints and the almost all-comprehensive book \cite{Hori:2003ic}. Even though it is not yet clear what mirror symmetry really is, it is probably one of the best understood (topological) string theory dualities.

Another duality has been conjectured long ago by 't Hooft \cite{'tHooft:1973jz} in the context of QCD-like theories for strong interactions and is known as large-$N$ duality or open/closed string duality. It relates open strings/gauge theories on one side with closed strings on the other side. 
Recently large-$N$ duality has been conjectured to be related to geometric transitions \cite{Gopakumar:1998ki, Cachazo:2001jy, Cachazo:2001gh, Cachazo:2001sg}. Geometric transitions interpret the resummation of the open string sector of an open-closed string theory as a transition in the target space geometry, connecting two different components of a moduli space of Calabi-Yau threefolds. 
In the B-model this ends up in many examples as a relation between the special geometry of the moduli space of complex structures of a Calabi--Yau threefold, the Kodaira--Spencer theory of gravity, on the closed string side and random matrices on the open string side \cite{Dijkgraaf:2002fc, Dijkgraaf:2002vw, Dijkgraaf:2002dh}. Also, large-$N$ duality has led on the A-model side to the exact computation of Gromov--Witten invariants for any (non-compact) toric Calabi--Yau threefolds by gluing elementary trivalent graphs together, \cite{Aganagic:2003db}. In \cite{Li:2004uf} these results were placed on firmer mathematical ground by using the relative Gromov--Witten theory.
Large-$N$ transitions touch profoundly the very nature of a quantum spacetime, since spacetime itself becomes an emergent feature with respect to the original gauge/open string formulation. We will actually come back again to geometric transitions, and in some sense they will be a guiding idea for the present work. 

A last duality, known as S-duality, has led to somehow similar results for the A-model, relating Gromov--Witten theory to Donaldson--Thomas theory and random partitions, \cite{Iqbal:2003ds, Okounkov:2003sp, MNOP}. It is a nonperturbative duality from the target-space viewpoint.

\paragraph{} 
In this thesis we shall attempt to plot a course, touching on some of these dualities and trying to identify a possible building block for B-type topological strings on Calabi--Yau threefolds. 
In actual fact, we shall just take a first step in that direction.
In particular, we will consider the dynamics of topological B-branes on curves in local Calabi--Yau threefolds.
In the B-model, the category of topological branes is conjectured to be equivalent to the bounded derived category of coherent sheaves on the threefold. We shall restrict ourselves to the objects that can be represented as complex submanifolds with vector bundles on them and we will mostly discuss the case of complex dimension $1$.

To motivate our study, let us take a closer look at large-$N$ dualities, relating open strings to closed strings.
For a very clear introductory discussion of large-$N$ dualities and geometric transitions in the context of topological strings, see \cite{Marino:2004uf,Marino:2004eq} and the book \cite{Marino:2005sj}.
Closed strings are defined by using two-dimensional conformal field theories (CFTs) coupled to two-dimensional gravity. In particular, a geometric vacuum is specified by a conformal $\sigma$-model, i.e. a theory of maps $x: \Sigma_g \to X$, where $\Sigma_g$ is a Riemann surface of genus $g$ and $X$ is the target manifold. 
One can compute the free energies $F_g(t_i)$ at genus $g$ as correlation functions of the two-dimensional $\sigma$-model coupled to gravity, where $t_i$ are geometric data of the target space $X$. The partition function of the string theory is given in a genus expansion as the generating function
\begin{eqnarray}\label{eq:closedstring}
F(g_s; t_i) := \sum_{g=0}^{\infty} g_s^{2g-2} F_g(t_i) ,
\end{eqnarray}
where $g_s$ is the string coupling constant (sometimes indicated as $\hbar$, $\lambda$ or $\epsilon$).

On the other hand, open string theory is defined as a theory of maps from an ``open'' Riemann surface $\Sigma_{g,h}$ of genus $g$ and with $h$ boundaries to a target manifold $X$ with fixed boundary conditions for these maps. One can impose Dirichlet boundary conditions by fixing a submanifold $S$ where the open strings have to end, i.e. the map sends the boundary of $\Sigma_{g,h}$ to $S$. Also, Chan--Paton factors will induce a $U(N)$ gauge symmetry.
This is a configuration with $N$ D-branes ``wrapped'' around $S$.
The open string amplitudes $F_{g,h}$ can be arranged 
in a generating function as follows
\begin{eqnarray}\label{eq:openstring}
F(g_s, N) := \sum_{g=0}^{\infty} \sum_{h=1}^\infty g_s^{2g-2+h} N^h F_{g, h} . 
\end{eqnarray}
In this expression the sum over $h$ can be formally resummed. One introduces the {'t Hooft parameter} $t:=g_s N$ and rewrites the expression (\ref{eq:openstring}) as (\ref{eq:closedstring}) by defining 
\begin{eqnarray}
F_g(t) := \sum_{h=1}^{\infty} F_{g,h} t^h . \nonumber
\end{eqnarray}
In this way, one is describing an open string theory in the presence of $N$ branes formally using a closed string genus expansion. Of course, one would like to know whether this expansion can actually be obtained from a closed string theory.
The first example in which an open string theory on a given background with D-branes was conjectured to be equivalent to a closed string theory on another background is known as AdS/CFT correspondence \cite{Maldacena:1997re}, reviewed in \cite{Aharony:1999ti}. In this case, type IIB open/closed string theory in $\mathbb{R}^{10}$ with $N$ D3 branes on $\mathbb{R}^4 \subset \mathbb{R}^{10}$ is conjectured to be equivalent in some limit to type IIB closed string theory on $AdS_5 \times S^5$. In particular, the open string side in this limit turns out to be the conformal 
$\mathcal{N} = 4$ $U(N)$ super-Yang-Mills theory in 4 dimensions.

As indicated above, D-branes can be understood as extended sources. In type II string D-branes are sources of the Ramond--Ramond field.
In topological strings, one expects branes to be sources of the K\"ahler $(1,1)$-form $J$ for the A-model and of the holomorphic $(3,0)$-form $\Omega$ for the B-model. Using this idea, it is possible to give an ``intuitive'' explanation of the effect of branes on the geometry, see for example \cite{Neitzke:2004ni} for a discussion. Let us consider a holomorphic 2-cycle $\Sigma \subset X$ around which a B-brane can be wrapped and a 3-cycle $M$, which links the 2-cycle $\Sigma$. This means that $M = \partial S$ for a 4-cycle $S$ intersecting $\Sigma$ once, in the same way that two real curves can link one another in a space of real dimension 3.
$M$ is homologically trivial as a cycle in $X$ (but nontrivial in $X - \Sigma$) and
\begin{equation}
\int_M \Omega = 0 \nonumber
\end{equation}
since $\de \Omega = 0$. The effect of $N$ B-branes on $\Sigma \subset X$ is to ``generate a flux of $\Omega$ through M''
\begin{equation}
\int_M \Omega = N g_s =: t \nonumber
\end{equation}
This fact 
suggests a privileged role for B-branes of complex dimension 1, since there is no other field for the branes in different dimensions. 
The ``classical'' example of this phenomenon is known as conifold transition. In the topological string setting it has first been studied for the A-model in \cite{Gopakumar:1998ki}. In the ``S-dual'' B-model case, one starts on the open side with the total space of the holomorphic vector bundle
\begin{equation}
   \mathcal{O}_{\mathbb{P}^1}(-1) \oplus \mathcal{O}_{\mathbb{P}^1}(-1) \ .
\end{equation}
with $N$ B-type branes on the zero section of the bundle. This local Calabi--Yau threefold is known as the \emph{resolved conifold} since it can be obtained as the small resolution of the singular hypersurface in $\mathbb{C}^4$
\begin{equation} 
   u v - x y = 0
\end{equation}
called the \emph{conifold}, a cone over the quadric surface $\mathbb{P}^1 \times \mathbb{P}^1 \subset \mathbb{P}^3$.
 The large-$N$ closed dual of this geometry is conjectured to be the so-called \emph{deformed conifold}, given as a hypersuface in $\mathbb{C}^4$
\begin{equation} 
   u v - x y = t
\end{equation}
where the parameter of the deformation $t \in \mathbb{C}$ is the 't Hooft parameter $N g_s$. In this example the cycles linking each other are topologically $S^2$ and $S^3$: when the first is homologically trivial the other is not (and vice versa). Note also that these spaces have the same ``asymptotic topology''. 
All the other known cases of large-$N$ dual Calabi-Yau threefolds in topological strings \cite{Dijkgraaf:2002fc, Dijkgraaf:2002vw, Dijkgraaf:2002dh, Ferrari:2003vp, Diaconescu:2005jv, Diaconescu:2005jw} are given as extremal transitions, which is very much in the spirit of the previous example.

\paragraph{}
With these motivations in mind, we will consider the class of noncompact smooth Calabi--Yau threefolds that are topologically fibrations over a compact Riemann surface, the \emph{local curves}, and study the dynamics of B-branes wrapped around the curve. There are different but closely related possible approaches to this problem. The first one is of course the worldsheet approach, but we will not consider it here. A second one, is the study of the open string field theory of the B-brane. Finally, the classical $g_s = 0$ limit of these dynamics for one single brane reduces to the deformation theory of the curve inside the Calabi--Yau threefold \cite{Bershadsky:1995qy}.

In chapter \ref{ch:localCY} we define the classes of
smooth Calabi--Yau threefolds that can be obtained as 
bundles over smooth curves and smooth surfaces with linear fibres, 
the \emph{local curves} and \emph{local surfaces}, and 
derive some of their properties. 

In chapter \ref{ch:sft} we consider the string field theory approach
to brane dynamics. 
The open string side of B-model has been shown to be represented by 
a six-dimensional holomorphic Chern--Simons theory \cite{Witten:1992fb}. 
The dynamics for lower dimensional branes can be obtained by 
dimensional reduction of the Chern-Simons theory to the world-volume of the brane. 
We give a reduction prescription for the holomorphic Chern--Simons action functional, 
following \cite{Bonelli:2005dc, Bonelli:2005gt}. 
We consider as target space of the theory a local curve, that is,
a $\mathbb{C}^2$ fibre bundle on a curve of given genus, with $N$ B-branes on the curve. 
After reduction to the branes one ends up with a holomorphically deformed 
$\beta$-$\gamma$ system on the curve, with the coupling constants of the 
deformation given by the complex structure of the target space. 
In some cases, this theory can be shown to reduce to a matrix model, 
suggesting some form of integrability for the original topological string theory.
This construction corresponds, in the framework of type IIB string theory, 
to spacetime filling D5-branes wrapped around two-dimensional cycles 
and to the engineering of a 4D effective gauge theory. 
The number of chiral multiplets in the adjoint representation of the gauge group 
is equal to the number of independent holomorphic sections of the normal bundle to the curve. 
The superpotential of the gauge theory is given by the genus zero topological amplitude.
From a similar construction \cite{Bonelli:2006nk}, one can also obtain 
chiral multiplets in the fundamental representation of the gauge group, 
the flavour fields.
The previous analysis should of course be compared with the calculations 
done using the derived category of coherent sheaves on the target space, 
and with the methods of \cite{Aspinwall:2004bs}.

In chapter \ref{ch:deformations} we consider the classical limit of the previous approach and the deformation theory of the curve in the threefold for a class of Calabi--Yau threefolds, the \emph{Laufer curves}. From a purely geometric viewpoint, the critical points of the D-brane superpotential are connected with the versal deformation space of the curve in the target Calabi-Yau space. This remark, together with the Serre duality between deformation space and obstruction space for curves in Calabi-Yau threefolds, suggested the existence of a single holomorphic function whose differential gives the obstruction map of the theory \cite{Katz:2000ab, Clemens:2002}, and whose derivatives should give the successive jets to the curve.
One wants to explicitely construct the relevant superpotential 
from the parameters of the geometry. 
In particular, following \cite{Bruzzo:2005dp}, we prove a conjecture 
by F. Ferrari \cite{Ferrari:2003vp} for the rational case, 
connecting the Hessian of the superpotential at a critical point 
to the normal bundle to the curve.
In this case, one also finds that the superpotential is connected 
to the geometric potential by a map that is dual to the 
``multiplication of the sections'' map, a result that is 
generalizable to the non-rational case \cite{Bruzzo:WIP}.


Thus, a more difficult problem, which also serves as a motivation for the previous work, is to understand more general large-$N$ transitions, for example in the case of the local curve in higher genus, with the hope of describing them as a renormalization group flow of the $6$-dimensional topological field theory.
For the local curve probably only some part of the moduli space admits a dual geometric description, and this should be related to asymptotic freedom in the physical theory. This is left to further investigation, and we just add a few more comments in the conclusions.

In the appendices we summarize the relation between topological strings and type II string theories and we give a short derivation of the open string field theory for the B-model with holomorphic Chern--Simons theory. We also recall some definitions of the objects used in the text. These appendices are simply intended as a list of useful background notions.

%

\newpage
$\ $ 
\pagestyle{empty}
\newpage
\thispagestyle{empty}

\pagenumbering{arabic}
\setcounter{page}{1}
\pagestyle{fancy}

\chapter{Local Calabi--Yau threefolds}
\label{ch:localCY}

In this chapter we introduce the geometries we shall use in the following chapters
and give some of their properties.
We consider noncompact smooth Calabi--Yau threefolds that can be obtained
as total spaces of (not necessarily holomorphic) fibre bundles with linear fibres over 
compact complex manifolds. We will call them \emph{local curves} 
when the base manifold has complex dimension $1$ and \emph{local surfaces} 
when it has complex dimension $2$.
Just for completeness, we could even define a \emph{local point} to be isomorphic to $\mathbb{C}^3$ and a \emph{local threefold} to be a compact Calabi--Yau threefold.

We consider these threefolds as tubular neighborhoods of smooth submanifolds of a
Calabi--Yau threefold, but we shall not discuss the convergence issues 
that arise when considering power series. These geometries could be understood
as formal schemes, see \cite{Hartshorne}.

\section{Local curves}


\begin{defin}\label{defin:localcurve}
Let $X$ be a complex threefold, $\dim_\mathbb{C} X = 3$, 
and $\Sigma$ a compact Riemann surface. Then $(X,\Sigma)$ is called
a \emph{(Calabi--Yau) local curve} if
\begin{enumerate}
\item $X$ is the total space of a $\mathbb{C}^2$ fibre-bundle over $\Sigma$
      as a differential manifold (\emph{local curve condition});
\item $X$ has trivial canonical bundle, $K_X \simeq \mathcal{O}_X$ 
      (\emph{Calabi--Yau condition});
\item $\Sigma$ is a complex submanifold of $X$.
\end{enumerate}
\end{defin}

Note that at this point we are not assuming $X$ to be a holomorphic bundle over 
$\Sigma$, that is, we are not assuming the existence of a holomorphic projection 
$\pi: X \to \Sigma$. 
Also, in general $\Sigma$ is not embedded as the zero section of the bundle.

We now consider more and more general examples of local curves,
starting with the simplest one, namely that of a rank-$2$ holomorphic vector bundle
on a compact Riemann surface, the \emph{linear local curve},
and then considering classes of nonlinear deformations.
For each example, we will first provide some general consideration 
and then the explicit transition functions of the threefold.
\begin{figure}\label{fig:localcurve}
\begin{center}
\includegraphics[height=3cm]{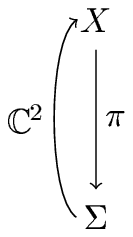}\hspace{1cm}
\includegraphics[width=6cm,height=4cm]{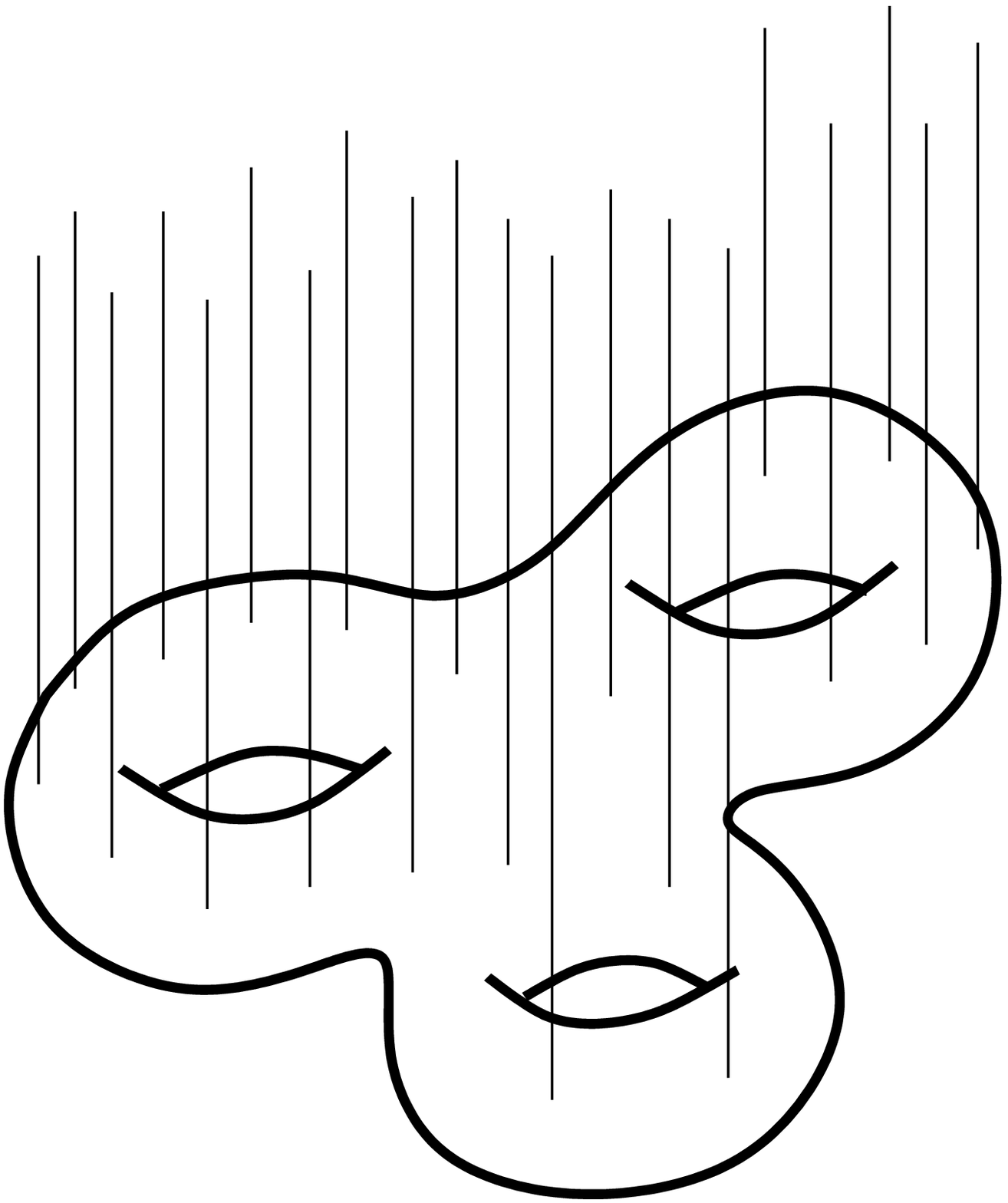}
\caption{The local curve, see the definition \ref{defin:localcurve}
         and the transition functions (\ref{eq:generallocalcurve}).
         The projection $\pi: X \to \Sigma$ is a holomorphic map only 
         when the terms $\{ \Psi^0 \}$ vanish.
         }
\end{center}
\end{figure}

\subsection{The linear local curve}

Let $\pi: V \to \Sigma$ be a holomorphic rank-$2$ vector bundle on a compact Riemann 
surface, see \cite{Gunning2}. Then there exist line bundles $\phi$ and $\phi'$ over $\Sigma$
such that $V$ is an extension of $\phi'$ by $\phi$, that is, there is given an exact 
sequence of analytic sheaves of the form
\begin{eqnarray}
  0 \to \phi' \to V \to \phi \to 0
\end{eqnarray}
where here and in the following we use the same symbol for the vector bundle and for
the sheaf of its sections. Extentions of one line bundle by another are classified
by the cohomology group $H^1 (\Sigma, \phi' \phi^{-1})$. Note that this group does not
give a classification of vector bundles up to isomorphisms. 
The trivial extension $\phi \oplus \phi'$ 
corresponds to the zero element of the group. 

The Calabi--Yau condition is equivalent to the condition
\begin{eqnarray}
  \det V \simeq K_\Sigma .
\end{eqnarray}
This can be seen for example using the sequence of vector bundles on $V$ 
(\emph{Atiyah sequence})
\begin{eqnarray}\label{eq:atiyahsequence}
  0 \to \pi^* V \to T_V \to \pi^* T_\Sigma \to 0
\end{eqnarray}
and
\begin{eqnarray}
  K_V \simeq (\det T_V)^{-1} \simeq \pi^* \left( (\det V)^{-1} K_\Sigma \right) .
\end{eqnarray}
As for the line bundles $\phi$ and $\phi'$, this implies $\phi' \simeq K_\Sigma \phi^{-1}$. 
Using Serre duality, one obtains the canonical isomorphism 
$H^1 (\Sigma, V) \simeq H^0 (\Sigma, V)^*$.

\paragraph{}
Let $\mathcal{U}=\{U_\alpha\}$ be an atlas for $\Sigma$ over which $V$ trivializes, $\{f_{\alpha \beta}(z_\beta)\}$ transition functions on $\Sigma$ for the given atlas, 
and $\{ \phi_{\alpha\beta} (z_\beta)\}$ and $\{\phi'_{\alpha\beta} (z_\beta)\}$ 
the transition functions for the line bundles $\phi$ and $\phi'$. 
Then, the transition functions for the vector bundle $V$ are
\begin{eqnarray}\label{eq:linearfibration}
\left\{\begin{array}{rcl}
   z_{\alpha}  &=& f_{\alpha \beta} (z_\beta)  \\ 
   \omega^1_{\alpha} &=& \phi_{\alpha\beta} (z_\beta) \,  \omega^1_{\beta}  \\
   \omega^2_{\alpha} &=& \phi'_{\alpha\beta} (z_\beta) \left( \omega^2_{\beta} + 
                                       2 \sigma_{\alpha \beta} (z_\beta) \omega^1_{\beta} \right) 
\end{array}\right.
\end{eqnarray}
where the family of functions $\{ \sigma_{\alpha \beta} \}$ gives a cocycle in
$Z^1 ( \mathcal{U}, \phi' \phi^{-1} )$. 
The Calabi--Yau condition in local coordinates reads
\begin{eqnarray}
  \phi'_{\alpha\beta} (z_\beta) = \phi_{\alpha\beta} (z_\beta) \left( f'_{\alpha \beta} (z_\beta) \right)^{-1} 
\end{eqnarray}
and the prime $'$ on the right hand side stands for derivation with respect to $z_\beta$.

\paragraph{}
Notice that vector bundles on a compact Riemann surfaces of genus $g$ are topologically classified by their rank and their degree, 
see for example \cite{LePotier}.
In our case we have a rank-2 vector bundle of degree $\deg (V) = \deg (K_\Sigma) = 2g - 2 \in \mathbb{Z}$. 
Thus, for a fixed genus $g$, all these vector bundles are 
isomorphic each other as differential manifolds. In particular 
they are all isomorphic to $T^* \Sigma \times \mathbb{R}^2$ .

\paragraph{}
In the rational case, i.e. $\Sigma \simeq \mathbb{P}^1$, by Grothendieck's
classification of vector bundles on curves of genus zero \cite{Grothendieck}, 
$V$ splits as a direct sum of two line bundles. The Calabi-Yau condition gives
\begin{eqnarray}
  V \simeq \mathcal{O}_{\mathbb{P}^1} (n) \oplus \mathcal{O}_{\mathbb{P}^1} (-n-2) , 
  \qquad n \in \mathbb{Z} , \ n \geq -1 .
\end{eqnarray}
In the standard atlas $\mathcal{U} = \{ U, U' \}$ of $\mathbb{P}^1$, 
the transition functions for the total spaces of these vector bundles read
\begin{equation}\label{eq:linearrationalfibration}
  \left\{\begin{array}{rcl}   
    z'  &=& 1/z  \\
    \omega_1' &=& z^{-n}  \omega_1   \\
    \omega_2' &=& z^{n+2} \omega_2  .
  \end{array}\right.
\end{equation}
Let us note \emph{en passant} that, among the local curves, 
those are the only examples of toric 
Calabi--Yau threefolds, since $\mathbb{P}^1$ is the only 1-dimensional 
complex compact toric manifold (i.e. 1-dimensional nonsingular complete toric variety). 
For the definition and the properties of toric varieties, see \cite{Fulton}
(and also \cite{Audin}, for the symplectic viewpoint). 

\paragraph{}
We want to consider now the space $H^1(V, T_V)$ of infinitesimal 
complex structure deformations of $V$.
At the risk of being tautological, these deformations correspond 
to vertices of the string theory, in particular to variations of
the string background. 
Our aim is to write the infinitesimal complex structure deformations of $V$ 
as elements of cohomology groups defined on the base $\Sigma$.
For this reason, let us first remark that the Atiyah sequence (\ref{eq:atiyahsequence}) 
induces a long exact sequence in cohomology of the form
\begin{eqnarray}
0 \to H^0(V, \pi^*V) \to H^0(V,T_V) \to H^0(V, \pi^* T_\Sigma) \to  \nonumber \\
 \to H^1(V, \pi^*V) \to H^1(V,T_V) \to H^1(V, \pi^* T_\Sigma) \to \cdots
\end{eqnarray}
Since $\pi$ is an affine map, we can use the proposition \ref{prop:affine}
in the appendix \ref{appendix-sheaves}
to reduce to cohomology groups on the base $\Sigma$ as follows
\begin{eqnarray}
0 \to H^0(\Sigma, \pi_*\pi^* V) \to H^0(V,T_V) \to H^0(\Sigma, \pi_*\pi^* T_\Sigma) \to  \nonumber \\
 \to H^1(\Sigma, \pi_*\pi^* V) \to H^1(V,T_V) \to H^1(\Sigma, \pi_*\pi^* T_\Sigma) \to 0
\end{eqnarray}
We now use the projection formula $\pi_*\pi^* V = V \otimes \pi_*\mathcal{O}_V$ (proposition \ref{prop:projection}) to obtain
\begin{eqnarray}
0 \to H^0(\Sigma, V \otimes \pi_*\mathcal{O}_V) \to H^0(V,T_V) \to H^0(\Sigma,  T_\Sigma \otimes \pi_*\mathcal{O}_ {T\Sigma}) \to  \nonumber \\
 \to H^1(\Sigma, V \otimes \pi_*\mathcal{O}_V) \to H^1(V,T_V) \to H^1(\Sigma,  T_\Sigma \otimes \pi_*\mathcal{O}_ {T\Sigma}) \to 0
\end{eqnarray}
Using $\pi_*\mathcal{O}_V = \bigoplus_{l\geq0} \Sym^l V^{*}$, we finally have the sequence
\begin{eqnarray}
0 \to \bigoplus_{l\geq0} H^0(\Sigma, V \otimes  \Sym^l V^{*}) \to H^0(V,T_V) 
       \to \bigoplus_{l\geq0} H^0(\Sigma,  K_\Sigma^{l-1}) \to  \nonumber \\
 \to \bigoplus_{l\geq0} H^1(\Sigma, V \otimes  \Sym^l V^{*}) \to H^1(V,T_V) 
       \to \bigoplus_{l\geq0} H^1(\Sigma,  K_\Sigma^{l-1}) \to 0 . 
\end{eqnarray}
Among these deformations, it is possible to recognize the group $H^1 (\Sigma,  T_\Sigma)$
of infinitesimal deformations of the base (first on the right) and the group
$H^1 (\Sigma,  \End\; (V))$
of infinitesimal deformations of the linear bundle structure (second on the left).
Notice also that all the deformations on the left, for a fixed Riemann surface $\Sigma$,
strongly depend on the vector bundle $V$; 
in fact, there is no reason to expect a smooth moduli space. 
On the other hand, the deformations on the right do not depend on the vector 
bundle $V$: they are the deformations ``along the base'' and do not preserve
the existence of the holomorphic projection $\pi: V \to \Sigma$.

Let us consider as an example the vector bundles on rational curves 
\begin{eqnarray}
V \simeq V_h := \mathcal{O}_{\mathbb{P}^1} (h-1) \oplus \mathcal{O}_{\mathbb{P}^1} (-h-1), \qquad h \geq 0
\end{eqnarray}
where $h:= \dim H^{0} ( \mathbb{P}^1, \mathcal{O}(n) ) = n+1$ for $n \geq -1$. In this case
\begin{eqnarray}
  V_h \otimes \Sym^{l} V_h^* \simeq \left( \mathcal{O}(h) \oplus \mathcal{O}(-h) \right) 
                                    \bigoplus_{m = 0}^l \mathcal{O} 
                                    \left( (l-2m)h +l-1 \right) .
\end{eqnarray}
For the \emph{resolved conifold}, i.e. $h=0$, 
the only deformations are those along the base, 
since the conifold has no nontrivial deformations. 
For $h=1$, we have that
\begin{eqnarray}
  \dim H^1 \left( \mathbb{P}^1, V_1 \otimes \Sym^{l} V_1^* \right) = 1 , \qquad l \geq 0 .
\end{eqnarray}

Let us remark that, among these complex structure deformations, 
we are interested only in those preserving 
the Calabi--Yau condition, that is, the triviality of the canonical bundle.

\subsection{The Laufer curve}\label{sec:Laufercurves}

The second example we consider is the Laufer curve.%
\footnote{The examples in \cite{laufer} are actually rational curves, 
here we consider their generalization in any genus.}
We shall define this class of local Calabi-Yau threefolds 
by giving their transition functions.

Let again $\Sigma$ be a smooth algebraic curve, 
$\mathcal{U}=\{U_\alpha\}$ an atlas for $\Sigma$ and $\{f_{\alpha \beta}(z_\beta)\}$
transition functions on $\Sigma$ for the given atlas. 
Let $\phi$, $\phi'$ two holomorphic line bundles on $\Sigma$ with transition
functions $\{ \phi_{\alpha\beta} (z_\beta)\}$ and $\{\phi'_{\alpha\beta} (z_\beta)\}$. 
We want to consider a nonlinear deformation 
of the vector bundle $\phi \oplus \phi'$ of the form
\begin{eqnarray}\label{eq:lauferfibration}
\left\{\begin{array}{rcl}
   z_{\alpha}  &=& f_{\alpha \beta} (z_\beta)  \\ 
   \omega^1_{\alpha} &=& \phi_{\alpha\beta} (z_\beta) \,  \omega^1_{\beta}  \\
   \omega^2_{\alpha} &=& \phi'_{\alpha\beta} (z_\beta) \left( \omega^2_{\beta} + \partial_{\omega^1_{\beta}} B_{\alpha\beta}(z_\beta, \omega^1_{\beta}) \right) .
\end{array}\right.
\end{eqnarray}
These transition functions define a bundle $X \to \Sigma$ with fibre $\mathbb{C}^2$. We will call  $\{B_{\alpha\beta}(z_\beta, \omega^1_{\beta})\}$ the \emph{geometric potential}, with $B_{\alpha\beta}$ holomorphic on $U_{\alpha\beta} \times \mathbb{C}$.

If we expand the geometric potential in its second variable
\begin{eqnarray}
 B_{\alpha\beta}(z_\beta, \omega^1_{\beta}) = \sum_{d=1}^{\infty} \sigma^{(d)}_{\alpha\beta}(z_\beta)  (\omega^1_{\beta})^d
\end{eqnarray}
then we have the following.

\begin{lemma}
Each coefficient $\sigma^{(d)}$ may be regarded as a cocycle defining an element in $H^1(\Sigma, \phi' \phi^{-(d-1)})$. 
\end{lemma}
\begin{proof}
The compatibility on triple intersections gives
\begin{eqnarray}
\omega^2_{\alpha} &=& \phi'_{\alpha\beta} (z_\beta) \left( \omega^2_{\beta} + 
                      d \sigma^{(d)}_{\alpha\beta}(z_\beta) (\omega^1_{\beta})^{d-1} \right) \nonumber \\
                  &=& \phi'_{\alpha\beta} (z_\beta) \left( \phi'_{\beta\gamma} (z_\gamma) \omega^2_{\gamma} + 
                      \phi'_{\beta\gamma} (z_\gamma)  d \sigma^{(d)}_{\beta\gamma}(z_\gamma) (\omega^1_{\gamma})^{d-1} + 
                                           \right. \nonumber \\
                  && \qquad \qquad                   \left. d \sigma^{(d)}_{\alpha\beta}(z_\beta) (\phi_{\beta\gamma} (z_\gamma))^{d-1} (\omega^1_{\gamma})^{d-1}
                                                     \right) \nonumber
\end{eqnarray}
Hence, we obtain for $\sigma^{(d)}$ the cocycle condition
\begin{eqnarray}
   \sigma^{(d)}_{\beta\gamma}(z_\gamma) +  \sigma^{(d)}_{\alpha\beta}(z_\beta) (\phi'_{\beta\gamma} (z_\gamma))^{-1} (\phi_{\beta\gamma} (z_\gamma))^{d-1} = 
   \sigma^{(d)}_{\alpha\gamma}(z_\gamma) \nonumber
\end{eqnarray}
We impose the following equivalence conditions on $\sigma$ and $\sigma'$. We consider the 
change of coordinates
\begin{eqnarray}
   \widetilde{\omega}^1_{\alpha} &=& \omega^1_{\alpha} \nonumber \\
   \widetilde{\omega}^2_{\alpha} &=& \omega^2_{\alpha} + h_\alpha(z_\alpha) (\omega^1_{\alpha})^{d-1}
   \nonumber
\end{eqnarray}
from which we obtain the coboundary condition
\[
    \sigma^{(d)}_{\alpha\beta}(z_\beta) -  \widetilde{\sigma}^{(d)}_{\alpha\beta}(z_\beta) = h_\beta(z_\beta) - h_\alpha(z_\beta) \ .  \qedhere
\]
\end{proof}

If we now impose the Calabi-Yau condition, i.e. triviality of the canonical bundle, we obtain that 
$\phi' \simeq K_\Sigma \phi^{-1}$, where $K_\Sigma$ is the canonical bundle on the curve,
and no condition on the geometric potential $B$. 
The cohomology groups of the previous lemma become 
$H^1(\Sigma, \phi' \phi^{-(d-1)}) = H^1(\Sigma, K_\Sigma \phi^{-d}) \simeq H^0(\Sigma, \phi^{d})^{*}$, 
where the last equality is Serre duality.

\begin{defin}
A \emph{Laufer (local) curve} is a noncompact Calabi--Yau threefold
defined by transition functions of the form (\ref{eq:lauferfibration}).
\end{defin}

In chapter \ref{ch:deformations} we will also assume that 
$H^0(\Sigma, \phi) := h > 0$ and $H^1(\Sigma, \phi) = 0$. 
%
Notice also that a Laufer curve can be seen as a kind of nonlinear
generalization of an extension of two line bundles.
In fact, it is a fibre bundle over a compact Riemann surface and 
moreover, given a holomorphic section $\Sigma$, 
there exist line bundles $\phi$ and $\phi'$ over $\Sigma$
and holomorphic maps 
\begin{eqnarray}
  &p:& X \to S_\phi \nonumber \\
  &i:& S_{\phi'} \to X
\end{eqnarray}
where $S_\phi$ and $S_{\phi'}$ are the total spaces of $\phi$ and $\phi'$.

\subsection{The holomorphic bundle case}
We now consider a more general case,
that is, we consider $\pi: X \to \Sigma$ a holomorphic $\mathbb{C}^2$ fibre bundle on 
a compact Riemann surface $\Sigma$. 
We can write the transition functions of $X$ as 
\begin{eqnarray}\label{eq:holomorphicfibration}
\left\{\begin{array}{rcl}
   z_{\alpha}  &=& f_{\alpha \beta} (z_\beta)  \\ 
   \omega^1_{\alpha} &=& \phi_{\alpha\beta} (z_\beta) w^1_{\alpha\beta} \left( z_\beta, \omega_{\beta}\right) \\
   \omega^2_{\alpha} &=& (f'_{\alpha\beta} (z_\beta))^{-1} (\phi_{\alpha\beta} (z_\beta))^{-1}
                            w^2_{\alpha\beta} \left( z_\beta, \omega_{\beta}\right) .
\end{array}\right.
\end{eqnarray}
As before, we choosed an atlas $\mathcal{U}=\{U_\alpha\}$ for $\Sigma$ over which $X$ trivializes, 
and $\{f_{\alpha \beta}(z_\beta)\}$ are the transition functions on $\Sigma$ for the given atlas.
For later convenience, and without loss of generality, we put in evidence
the transition functions of two line bundles $\phi$ and $\phi' \simeq K_\Sigma \phi^{-1}$.
For each $z_\beta \in U_{\alpha \beta}$, the transition functions give invertible holomorphic maps 
\begin{eqnarray}
  w_{\alpha\beta} (z_\beta, \cdot ): \mathbb{C}^2 \to \mathbb{C}^2 .
\end{eqnarray}

The Calabi--Yau condition, i.e. the existence of a nowhere vanishing
holomorphic $(3,0)$-form $\Omega$, implies the transition functions 
\begin{eqnarray}
  \de z_\alpha \wedge \de \omega^1_\alpha \wedge \de \omega^2_\alpha = 
  \de z_\beta  \wedge \de \omega^1_\beta  \wedge \de \omega^2_\beta .
\end{eqnarray}
These last conditions in turn implies that
\begin{eqnarray}
  \de w^1_{\alpha\beta} \wedge \de w^2_{\alpha\beta} = \de \omega^1_\alpha \wedge \de \omega^2_\alpha
\end{eqnarray}
for each $z_\beta \in U_{\alpha\beta}$, i.e. the transition functions are holomorphic symplectic transformations
\begin{eqnarray}
  w_{\alpha \beta}: U_{\alpha\beta} \to \mathfrak{Sp}(\mathbb{C}^2) .
\end{eqnarray}
Thus, the solution of the Calabi--Yau condition can be given in terms of a
set of potential functions $\chi_{\alpha\beta}$ (one for each double patch intersection modulo
triple intersections identities), the \emph{geometric potential}, 
which generates the transition functions as 
\begin{eqnarray}
  \epsilon_{ij} w^i_{\alpha\beta} \de w^j_{\alpha\beta} =
  \epsilon_{ij} \omega^i_{\alpha} \de \omega^j_{\alpha} - \de \chi_{\alpha\beta} .
\end{eqnarray}

If we consider the transition functions of $X$ as deformations
of a linear bundle and write
\begin{eqnarray}
w^i_{\alpha\beta}=\omega^i_{\alpha}+ \Psi^i_{\alpha\beta}(z_{\beta},\omega_{\beta}) ,
\end{eqnarray}
then the deformed complex structure preserves the Calabi--Yau condition if
in any $U_{\alpha} \cap U_{\beta}$ we have $\det\left(1+\partial\Psi\right)=1$, 
where $(1+\partial\Psi)^i_j=\delta^i_j+ \partial_j\Psi^i$.

Let us specify the previous construction in the case $\Sigma={\mathbb P}^1$. 
In this case the generic variation is 
\begin{equation}\label{variation}
\omega'_i=z^{n_i} \left( \omega^i + \Psi^i \left(z, \omega \right) \right)
\end{equation}
where $n_1:=n$ and $n_2:=-n-2$.
The functions $\Psi^i$ in (\ref{variation}) are analytic on ${\mathbb C}^* \times {\mathbb C}^2$, 
that is, they are allowed to have poles of finite order at $0$ and $\infty$ in $z$ 
and have to be analytic in $\omega=(\omega_1, \omega_2)$. 
The Calabi-Yau condition is solved by a single
potential function as follows
\begin{equation}\label{variation2}
\epsilon_{ij}w^id w^j = \epsilon_{ij}\omega^id \omega^j - \de \chi,
\end{equation}
where, as before, $w^i=\omega^i+\Psi^i(z_S,\omega)$.

\subsection{The general deformations}

A general complex structure deformation 
is given by the transition functions
\begin{equation}\label{eq:generallocalcurve}
\left\{\begin{array}{rcl}
z_{\alpha} &=& f_{\alpha\beta}(z_{\beta}) 
               \left( 1 + \Psi^0_{\alpha\beta} \left(z_{\beta},\omega_{\beta}\right) \right) \\
\omega^i_{\alpha} &=& V^i_{j\alpha\beta} \left(z_{\beta} \right)
                      \left( \omega^j_{\beta} + \Psi^j_{\alpha\beta}
		      \left(z_{\beta},\omega_{\beta}\right) \right) \ , 
		      \quad i,j=1,2\0
\end{array} \right.
\end{equation}
where $\{  V_{\alpha\beta} \}$ defines a rank-2 holomorphic vector bundle on $\Sigma$ and
$\Psi^i_{\alpha\beta}$ are analytic functions on double patch intersections, constrained by 
the chain rules on multiple patch intersections. 
The deformation is trivial if it can be reabsorbed via an analytic change of 
coordinates.

The Calabi--Yau condition reads
\begin{eqnarray}
  \de z_\alpha \wedge \de \omega^1_\alpha \wedge \de \omega^2_\alpha = 
  \de z_\beta  \wedge \de \omega^1_\beta  \wedge \de \omega^2_\beta .
\end{eqnarray}

\section{Local surfaces}

Most of the considerations of the previous section could be extended also to 
the case of local surfaces. In the following we briefly consider a few examples of
the linear case.

Let $S$ be a two-dimensional complex manifold. 
The total space $X_S$ of a line bundle on $S$ is a Calabi--Yau threefold if 
the line bundle is the canonical bundle of the surface $S$.
The complex manifold is defined by the transition functions
\begin{equation}\label{eq:linearsurface}
\left\{\begin{array}{rcl}
  {z}_{\alpha} &=& {f}_{\alpha\beta}\left({z}_{\beta}\right) \\
  p_{\alpha} &=& [\det {X}_{\alpha\beta}]^{-1} p_{\beta},
          \quad {\rm where} \quad 
          [{X}_{\alpha\beta}]:= \partial_{{z}_{\beta}}{f}_{\alpha\beta} \\
\end{array} \right.
\end{equation}
where $\{ U_{\alpha} \}$ is an atlas on $S$ extending to an atlas on $X_S$ 
by $W_{\alpha} = U_{\alpha} \times \mathbb{C}$ and
$\mathbf{z} = (z^1,z^2)$ are the local coordinates of $S$. 

The only toric local surfaces are those constructed from the projective plane
$\mathbb{P}^2$, from the Hirzebruch surfaces $F_a$ (the total spaces of the projective 
bundles $\mathbb{P}(\mathcal{O}(a) \oplus \mathcal{O}) \to \mathbb{P}^1$) and from 
successions of blow-ups of $\mathbb{P}^2$ or $F_a$ at fixed points under the toric action, 
see \cite{Fulton}.


\chapter[Topological string fields on the local curve]{Topological string fields\\ on the local curve}
\label{ch:sft}

In this chapter we describe the reduction of the topological type-B open string 
field theory to holomorphic cycles in local Calabi-Yau threefolds 
\cite{Bonelli:2005dc,Bonelli:2005gt}, that gives the dynamics of topological
branes on these cycles. In particular cases and for branes on rational curves,
the field theory on the brane further reduces to a multi-matrix model with couplings connected
with the complex structure parameters of the target space.
Topological type-B branes on holomorphic two-cycles of a Calabi-Yau local curve
(see the previous chapter) is what we call the B-side of the local curve.
For a very recent discussion of related topics, see \cite{Marino:2006hs}, and for the A-side of the local curve, see for example \cite{Caporaso:2006gk} and the references therein. 

These computations are strictly related to the properties of the ``physical'' (type II)
theory compactified on backgrounds of the form $\mathbb{R}^{1,3} \times X$, 
where $X$ is a Calabi-Yau threefold. 
In the presence of BPS branes and fluxes, the theory generically 
produces low energy effective theories with $\mathcal{N}=1$ supersymmetry. The topological 
B-model computes particular topological terms of the low energy effective action 
(see the appendix \ref{appendix-typeII} for further details). 
While the relation between open/closed string moduli and effective gauge 
theories is quite well understood in the case of $\mathcal{N}=2$ supersymmetry, 
the $\mathcal{N}=1$ case still lacks of a complete understanding.
For this reason, the study of the dynamics of branes in Calabi--Yau manifolds 
has attracted a lot of attention both in connection with its theoretical and phenomenological 
applications, e.g. \cite{Katz:1996th,Douglas:2000gi,Douglas:2001hw}. 
In particular, considering D-branes wrapped around two-cycles 
in a non-compact Calabi--Yau manifold one can link the superpotential of 
the $\mathcal{N}=1$ supersymmetric gauge theories living on the space-filling
branes to the deformation of the Calabi--Yau geometry \cite{Kachru:2000ih, Aspinwall:2004bs}. 
See also the discussion in chapter \ref{ch:deformations}.

The topological open B-model on a Calabi--Yau threefold $X$ with $N$ space-filling B-branes
(i.e. wrapping the whole Calabi--Yau threefold) 
 can be obtained from open string field theory and reduces 
\cite{Witten:1992fb} to the holomorphic Chern-Simons theory 
on $X$; %
we give a short derivation of this result in the appendix \ref{appendix-topologicalSFT}.
The action of holomorphic Chern-Simons theory is
\begin{eqnarray}\label{hCS}
S(\mathsf{A}) = \frac{1}{g_s} \int_X 
\Omega \wedge \Tr \left( \frac{1}{2}  \mathsf{A} \wedge \bar\partial \mathsf{A} + 
                         \frac{1}{3} \mathsf{A} \wedge \mathsf{A} \wedge \mathsf{A} \right)
\end{eqnarray}
where $\mathsf{A}$ is a $(0,1)$-form connection on $X$ with values in $\End(E)$, with 
$E$ a holomorphic vector bundle on $X$, and $\Omega$ is a holomorphic $(3,0)$-form on $X$.  
The theory described by this action is a very peculiar gauge theory in six real dimensions. 
Basically one could say that the theory is not really well defined, since the fuctional is not 
gauge invariant. Moreover it is a chiral (or holomorphic) theory and one should provide the ``slice'' in the space of fields over which one is integrating. We briefly discuss these topics in the appendix \ref{appendix-topologicalSFT}, section \ref{appendix-hCS}, but we shall not consider them again in the following sections.

The dynamics of a lower dimensional B-brane on a complex submanifold $S \subset X$ can be 
described by reducing the open string field theory from the space $X$ to the
B-brane world-volume $S$. 
It would be very interesting to obtain the action of
lower dimensional branes from a microscopic analysis.
In the language of derived categories, these branes correspond to 
complexes with just one non-trivial coherent sheaf supported on the submanifold, see for example
\cite{Aspinwall:2004jr} for details.

Since all the considerations are purely local, we can restrict ourself to non-compact
Calabi--Yau threefolds obtained as deformation of holomorphic vector bundles on the holomorphic cycle $S$,
called \emph{local curve} when $\dim_\mathbb{C} S=1$ and \emph{local surface} when $\dim_\mathbb{C} S=2$. Most of our considerations will be related to the case of the local curve, see figure \ref{fig:localcurve}.
Let $\Sigma$ be a compact Riemann surface and $\{ U_\alpha \}$ an atlas for $\Sigma$;
the transition functions for $X$ can be written
\begin{equation}\label{eq:localcurve}
\left\{\begin{array}{rcl}
z_{\alpha} &=& f_{\alpha\beta}(z_{\beta}) + \Psi^0_{\alpha\beta}
		      \left(z_{\beta},\omega_{\beta}\right) \\
\omega^i_{\alpha} &=& M^i_{j\alpha\beta}\left(z_{\beta} \right)
                      \left[ \omega^j_{\beta} + \Psi^j_{\alpha\beta}
		      \left(z_{\beta},\omega_{\beta}\right) \right]\ , 
		      \quad i,j=1,2\0
\end{array} \right.
\end{equation}
where $f_{\alpha\beta}$ are the transition functions on the base,
$M_{\alpha\beta}$ the transition functions of a holomorphic vector bundle $V$ 
and $\Psi^j_{\alpha\beta}$ are deformation terms, holomorphic on the 
intersections $(U_\alpha \cap U_\beta) \times \mathbb{C}^2$.

The Calabi-Yau condition on the space $X$, i.e. the existence of a nowhere
vanishing holomorphic top-form $\Omega = \de z \wedge \de w^1 \wedge \de w^2$, 
puts conditions on the vector bundle and on the deformation terms. The 
determinant of the vector bundle $V$ has to be equal to the canonical line bundle 
on $\Sigma$ and for the transition functions this means 
$\det M_{\alpha \beta} \times f'_{\alpha\beta}=1$. 
When $\Psi^0 = 0$, for the deformation 
terms we have $\det \left(1+\partial\Psi\right) = 1$, where 
$(1+\partial\Psi)^i_j=\delta^i_j+ \partial_j\Psi^i$.
The solution of this condition can be given in terms of a set of potential 
functions $\chi_{\alpha\beta}$, the \emph{geometric potential}, which generates 
the deformation via 
\begin{eqnarray}
\epsilon_{ij} w^i_{\alpha\beta} \de w^j_{\alpha\beta} = \epsilon_{ij} 
\omega^i_{\alpha} \de \omega^j_{\alpha} - \de \chi_{\alpha\beta}, \0
\end{eqnarray}
where we define the \emph{singular coordinates} 
$w^i_{\alpha\beta}= \omega^i_{\alpha} + \Psi^i_{\alpha\beta}(z^{\beta},\omega_{\beta})$.

In the first section we give a reduction prescription and 
obtain the reduced action for the linear geometry $\Psi^i \equiv 0$,
i.e. for a rank-2 holomorphic vector bundle $V \to \Sigma$,
restricting ourselves to the case in which $E$ is trivial.
The result turns out to be
\begin{eqnarray}
 S_{red} = \frac{1}{g_s} \int_\Sigma 
           \Tr \left( \frac{1}{2}\epsilon_{ij}\varphi^i D_{\bar z}\varphi^j \right) \nonumber
\end{eqnarray}
where $(\varphi^1, \varphi^2)$ are sections of $V \otimes \End (E)$
and $D_{\bar z}:= \bar\partial + [A_{\bar z}, \cdot]$.
We also discuss the reduction in the linear case for the local surface, obtaining a BF-like action.

In the second section we consider the reduction for the non-linear case.
In the abelian case the cubic term 
in the holomorphic Chern-Simons lagrangian is absent. 
The geometric potential gives the deformation of the action due to the deformation 
of the complex structure and one obtains for $\mathbb{P}^1$
\begin{eqnarray}
S_{red} = \frac{1}{g_s}
          \left\{ \frac{1}{2} \int_{\mathbb{P}^1}  \epsilon_{ij}\phi^i \partial_{\bar z} \phi^j 
          + \frac{1}{2}\oint_{\mathcal{C}_0} \frac{\de z}{2 \pi \ii} \chi (z,\phi) \right\} . \nonumber
\end{eqnarray}
Therefore the reduced theory is a $\beta$--$\gamma$ system with a junction interaction along the equator.
On a curve of genus $g \geq 1$, one finds similar results facing some
difficulty in writing the interaction term as an integral on cycles of the curve.
In the {non-abelian case} it is necessary to specify a matrix ordering. 
It is possible to avoid matrix ordering prescriptions if
$\chi(z,\omega)$ does not depend on one coordinate along the fibre, that is, for the Laufer rational curve.
Defining $B := \omega^1 \Psi^2 + \chi$ the reduced action reads
\begin{eqnarray}\label{Sred}
 S_{red} = \frac{1}{g_s} 
           \left\{ \int_{\mathbb{P}^1} \Tr(\phi^2 D_{\bar z} \phi^1) + 
             \oint_{\mathcal{C}_0}  \frac{\de z}{2 \pi \ii} \Tr B(z,\phi^1) \right\} .  
\end{eqnarray}
In this case, after gauge fixing, one finds the \emph{matrix model}
\begin{eqnarray}\label{functint}
Z_N(g_s, \{t\}) = \int \prod_{i=0}^n d X_i
e^{-\frac{1}{g_s}{W}\left(X_0,\dots,X_n\right)} \nonumber
\end{eqnarray}
with superpotential/matrix potential
\begin{eqnarray}
W \left(X_0,\dots,X_n\right) = \sum_{d=0}^{\infty} \sum_{k=0}^{dn} t^{(k)}_d \sum_{i_1, \dots, i_d=0 \atop i_1+ \dots + i_d=k }^n 
                                X_{i_1} \dots X_{i_d} \nonumber
\end{eqnarray}
and \emph{times} $t^{(k)}_d$ corresponding to the parameters of the complex structure deformation.
This result generalizes \cite{Kachru:2000ih,Dijkgraaf:2002fc,Dijkgraaf:2002vw,Dijkgraaf:2002dh} and confirms \cite{Ferrari:2003vp}. Also, it suggests some form of integrability for the underlying topological string theory.
In fact, matrix models have been extensively studied in the case of one matrix and of two matrices with bilinear interaction, see \cite{Bonelli:2005dc,Bonelli:2005gt,Bonelli:2006nk} and the references therein. The other cases like those we found from the topological B-model still lack a complete solution.

In the last section we give a reduction prescription that produces the matrix model connected 
to gauge theories with flavours. We also propose an interpretation in terms 
of a configuration of zero and one dimensional branes.


\section{The linear case}

\subsection{The linear local curve and the $\beta$-$\gamma$ system}

Let us consider a non compact Calabi--Yau threefolds that is the total space of a rank-$2$ 
holomorphic vector bundle $V \to \Sigma$, where $\Sigma$ is a compact Riemann surface,
i.e. a linear local curve.
Recall that, chosen a trivialization of $V$, 
the atlas $\{ U_{\alpha} \}$ on $\Sigma$ extends to an atlas on $V$ 
by $V_{\alpha} := U_{\alpha} \times {\mathbb C}^2$ and that
the complex manifold $V$ is defined by the transition functions
\begin{eqnarray}\label{eq:vectorbundle}
\left\{\begin{array}{rcl}
   z_{\alpha}  &=& f_{\alpha \beta} (z_\beta)  \\ 
   w^i_{\alpha} &=& M^{i}_{j \alpha\beta} (z_\beta) w^j_{\beta} \qquad i,j=1,2
\end{array}\right. 
\end{eqnarray}
in any double intersection $U_{\alpha\beta}:= U_{\alpha} \cap U_{\beta}$,
where $z_\alpha$ are local coordinates on $\Sigma$ 
and $w_\alpha=(w^1_\alpha, w^2_\alpha)$ on the ${\mathbb C}^2$ fibres.
Note also that the normal bundle to the curve and the vector bundle $V$
are isomorphic, $N_{\Sigma | V} \simeq V$. 
The embedding equations for $\Sigma$ in $V$ are $w_\alpha^i=0$, 
i.e. $\Sigma$ is embedded as the zero section of the bundle.
Requiring the complex manifold $V$ to be a Calabi--Yau threefold,
restricts the determinant bundle of $V$ to be equal to the canonical line bundle of $\Sigma$, 
$\det V \simeq K_\Sigma$. 
In local coordinates the Calabi--Yau condition reads
\begin{eqnarray}
( \det M_{\alpha\beta} ) \times f'_{\alpha\beta} = 1 . 
\end{eqnarray}
Under this condition, $V$ is equipped with a nowhere vanishing holomorphic 
$(3,0)$-form $\Omega$, that in local coordinates can be written as follows
\begin{eqnarray}
  \Omega |_{U} = \de z \wedge \de w^1 \wedge \de w^2 .
\end{eqnarray}
Here and henceforth we suppress the indices $\alpha$, $\beta$ 
when no confusion can arise.

Given a holomorphic rank-$N$ vector bundle $E \to V$,
the string field theory for the B-model with these $N$ topological B-branes on $V$
is the \emph{holomorphic Chern-Simons theory}, \cite{Witten:1992fb}. 
The action of the theory is
\begin{equation}\label{hCS2}
  S(\A)=\frac{1}{g_s} \int_{V} {\cal L}, \quad 
  \mathcal{L} = \Omega \wedge \Tr \left(\frac{1}{2} \A \wedge \bar\partial \A + 
                \frac{1}{3} \A \wedge \A \wedge \A \right)
\end{equation}
where $\mathsf{A}$ is a $(0,1)$-form connection on $V$ with values in $\End(E)$. 
The dynamics of B-branes wrapped around the holomorphic 2-cycle $\Sigma$
can be described by reducing the open string field theory from the total space $V$ 
to the B-brane worldvolume $\Sigma$.
Since the bundle $V$ is nontrivial, the reduction of the lagrangian
has to be prescribed patch by patch in such a way that the end product 
is independent of the chosen trivialization.

In the following we propose such a reduction prescription,
restricting ourselves to the case in which $E$ is trivial, 
$E \simeq V \times \mathbb{C}^N$. 
The lagrangian $\mathcal{L}$ in (\ref{hCS2}) is a $(3,3)$-form on $V$;
our purpose is to restrict it to a $(1,1)$-form $\mathcal{L}_{red}$ on $\Sigma$. 
Thus, first we split the form $\mathsf{A}$ into horizontal and vertical components 
using a reference connection $\Gamma$ on the vector bundle $V$ and we impose 
to these components independence to the vertical directions. 
We obtain a (3,3)-form on $V$ that is independent to the coordinates along the fibre.
Then, we obtain a $(1,1)$-form on $V$  contracting the $(3,3)$-form 
with a $(2,2)$-multivector; this multivector is constructed 
using a bilinear structure $K$ on the bundle $V$. 
If the connection $\Gamma$ is the generalized Chern connection for the 
bilinear structure $K$, then the result is independent of the chosen $(\Gamma, K)$.
Finally, we choose a section of $V$ and use it to pullback this form to the base $\Sigma$.
The $(1,1)$-form $\mathcal{L}_{red}$ obtained in this way
does not depend on the chosen section.

As for the first step, let us forget for a moment that
$\A$ is a $(0,1)$-form connection on $E$ and just consider 
(here we are using the fact that $E$ is a trivial bundle and also
restricting to gauge tranformations that are constant along
the fibres)
\begin{equation}
  \A \in \Omega^{(0,1)}_V .
\end{equation}
In order to impose indipendence of $\A$ along the fibre, we want to 
split $\A$ in horizontal and vertical components.
If we consider the exact sequence of vector bundles on $V$
(\emph{Atiyah sequence})
\begin{eqnarray}\label{eq:atiyahsequence2}
  0 \to \pi^* V \to T_V \to \pi^* T_\Sigma \to 0,
\end{eqnarray}
we can take its dual and its complex conjugate, to obtain
\begin{eqnarray}\label{eq:atiyahsequence3}
  0 \to \pi^* \Omega^{(0,1)}_\Sigma \to \Omega^{(0,1)}_V \to \pi^{*} \bar{V}^* \to 0 ,
\end{eqnarray}
where we denote by $V^*$ the dual vector bundle, glueing with $(M^{-1})^t$, 
and by $\bar V$ the complex conjugate one. 
If we now choose a connection on $V$ compatible with the holomorphic
structure, it will induce a connection on $\bar{V}^*$ compatible with
the (anti)holomorphic structure. With this connection we can produce 
a splitting of the above dualized Atiyah sequence. 

In local coordinates, this corresponds to the following procedure.
In any local chart, there is a local notion of parallel and 
transverse directions along which we split 
$\A = \A_{\bar{z}} \de \bar{z} + \A_{\bar{\imath}} \de  w^{\bar{\imath}}$.
The parallel part $\A_{\bar{z}}$ glues on double patches intersections as an 
invariant $(0,1)$-form connection on $\Sigma$ only when restricted
to the base, while otherwise gets also a linear contribution in
$w$ due to the generic non triviality of $V$.
The transverse coefficients $\A_{\bar{\imath}}$ glue
as a section of $\bar V^*$. 
Since the reduction to the base has to be performed covariantly, 
we have to expand $\A_{\bar{z}} = 
   A_{\bar{z}} - A_{\bar{k}} \Gamma_{\bar{z} \bar{\jmath}}^{\bar{k}} w^{\bar{\jmath}}$
and $\A_{\bar i}=A_{\bar i}$, where $A_{\bar z} \de \bar z \in \Omega^{(0,1)}_\Sigma$,
$A_{\bar i}\in \bar V^*$ and 
$\de \bar{z} \Gamma_{\bar{z}\bar{\jmath}}^{\bar k}$ is the induced $(0,1)$-form connection on $\bar V$.
The reduction process is defined by specifying the subfamily of 
connections we limit our consideration to. Our prescription is that the 
matrix valued dynamical fields $(A_{\bar z}, A_{\bar{\imath}})$ that survive the
reduction are those independent of the coordinates along ${\mathbb C}^2$.
A direct calculation from the lagrangian $\mathcal{L}$ in (\ref{hCS2})
for the above reduced configurations gives
\begin{equation}\label{redlag2}
L = \Omega \wedge \Tr \frac{1}{2} \left(
                      A_{\bar{\imath}} D_{\bar z} A_{\bar{\jmath}} + 
                      A_{\bar{\imath}} \Gamma_{\bar{z}\bar{\jmath}}^{\bar k} A_{\bar k}
                                        \right)
                      \de w^{\bar{\imath}} \wedge \de \bar{z} \wedge \de w^{\bar{\jmath}}
\end{equation}
where $D_{\bar z}$ is the covariant derivative with respect to the gauge structure.
Also notice that the above does not depend on the base representative, that is 
on the values of $w^i$.

To obtain a $(1,1)$-form on $\Sigma$ from the expression (\ref{redlag2}),
we have to couple the above reduction with the contraction 
of the differentials along the fibre directions to obtain a well defined 
$(1,1)$-form on $\Sigma$. 
To define this operation, let us consider a bilinear structure $K$ in $V$, 
that is a local section $K\in \Gamma( V \otimes \bar V)$, 
the components $K^{i \bar \jmath}$ being an invertible complex matrix at any point.

The derivation of the basic $(1,1)$-form is realized patch by patch 
with the help of $K$ as contraction of the 
(3,3)-form Lagrangian by the two bi-vector fields 
$k=\frac{1}{2}
\epsilon_{ij} K^{i \bar l} K^{j \bar k}
\frac{\partial}{\partial \bar w^l}
\frac{\partial}{\partial \bar w^k}$ 
and $\rho=\frac{1}{2} \epsilon^{ij}\frac{\partial}{\partial w^i}
\frac{\partial}{\partial w^j}$.
Notice that $k \in \det V$ and $\rho \in \det V^*=(\det V)^{-1}$
so that the combined application of the two is a globally well defined
operation.
Calculating then the pullback Lagrangian, we obtain
\begin{equation}\label{pb'}
\mathcal{L}_{red} = i_{\rho \wedge k} L=
\um \de z \de {\bar z} (\det K) 
\epsilon^{\bar \imath \bar \jmath}
\Tr \left[
A_{\bar \imath} D_{\bar z} A_{\bar \jmath} + 
A_{\bar \imath} \Gamma_{\bar z\bar \jmath}^{\bar k} A_{\bar k}
\right]
\end{equation}

Our last step relates the reference connection and the reference bilinear 
structure in order to obtain a result which is independent to the 
trivialization we used. 
Let us define the field components $\varphi^i=i_{V^i}A\in V$, 
where $V^i=K^{i \bar \jmath}\frac{\partial}{\partial\bar w^j}$
and plug it into (\ref{pb'}). One gets
\begin{equation}\label{pb''}
{\cal L}_{red}= 
\um \de z \de {\bar z}  
\Tr \left[\epsilon_{ij} \varphi^i D_{\bar z} \varphi^j + 
(\det K) \varphi^m \varphi^n \epsilon^{\bar \imath \bar \jmath}
\left(
K_{m \bar \imath} \partial_{\bar z} K_{n \bar \jmath}
+ K_{m \bar \imath} K_{n \bar k}
\Gamma_{\bar z\bar \jmath}^{\bar k}
\right)\right]
\end{equation}
where $K_{\bar \imath j}$ are the components of the inverse bilinear structure, 
that is $K_{\bar \imath j} K^{j \bar l}=\delta_{\bar \imath}^{\bar l}$.
In order to have a result which is independent of the trivialization, just
set the reference connection to be the ``generalized'' Chern connection of 
the bilinear structure $K$, that is
$\Gamma_{\bar z \bar \jmath}^{\bar k}= K_{\bar \jmath l}\partial_{\bar z} K^{l \bar k}$, 
where as we said $K$ is not necessarily hermitian.
Therefore, choosing our reference trivialization $(\Gamma, K)$ data 
to satisfy this natural condition, we get 
\begin{equation}
{\cal L}_{red}= \um
\de z \de {\bar z} \Tr \left[
\epsilon_{ij}\varphi^iD_{\bar z}\varphi^j 
\right]
\label{pb}\end{equation}
which is a well defined $(1,1)$-form on $\Sigma$. Hence the
action for the reduced theory is given by
\begin{eqnarray}
S_{red} = \frac{1}{g_s}\int_\Sigma {\cal L}_{red}=
\frac{1}{g_s}\um\int_\Sigma \de z \de{\bar z} \Tr\left[
\epsilon_{ij} \varphi^i D_{\bar z} \varphi^j 
\right] .
\end{eqnarray}

This calculation generalizes to the non-abelian case the argument of \cite{Aganagic:2000gs}.

\subsection{The linear local surface and the $B$-$F$ system}

In this section we extend the method for the reduction of the holomorphic Chern--Simons 
functional to the non compact Calabi--Yau geometry around a holomorphic four cycle $S$.
This geometry describing the cycle
is the total space of the canonical line bundle $K_S$, 
that is the bundle of the top holomorphic forms on $S$. We denote this space as $X_S=tot(K_S)$.

Any atlas $\{ U_{\alpha} \}$
on $S$ extends to an atlas on $X_S$ 
by $\hat U_{\alpha} = U_{\alpha} \times {\mathbb C}$.
The complex manifold is defined by the gluing map between the patches
\begin{equation}\label{eq:linear'}
\left\{\begin{array}{rcl}
{\bf z}_{\alpha} &=& {\bf f}_{\alpha\beta}\left({\bf z}_{\beta}\right) \\
p_{\alpha} &=& [\det {\bf X}_{\alpha\beta}]^{-1} p_{\beta}, \quad {\rm where} \quad
[{\bf X}_{\alpha\beta}]=
\partial_{{\bf z}_{\beta}}{\bf f}_{\alpha\beta} \\
\end{array} \right.
\end{equation}
in any double patch intersection $U_{\alpha} \cap U_{\beta}$.
In (\ref{eq:linear'}) and in the following, $\mathbf{z} = (z^1,z^2)$ denotes the two 
complex coordinates on $S$. 
The holomorphic $(3,0)$-form on $X_S$ is $\Omega = dz^1 \wedge dz^2 \wedge dp$.

Let us consider the topological B-model on 
$X_S$. In this case, D-branes can wrap the 4-cycle $S$ and
the theory describing the dynamics of these objects is obtained then by 
reducing the holomorphic Chern--Simons functional to the D-brane world-volume.
We consider here again only the case in which the gauge bundle $E$ is trivial.

The action of holomorphic Chern--Simons theory is
\begin{equation}
S(\A)=\frac{1}{g_s}\int_{X_S} {\cal L}, 
\quad
{\cal L}=
\Omega\wedge \Tr\left(\frac{1}{2} \A\wedge
\bar\partial \A + \frac{1}{3} \A \wedge \A \wedge \A\right)
\label{hCS'}\end{equation}
where $\A\in T^{(0,1)}\left(X_S\right)$.

We split $\A=\A_{\bar\z}d\bar\z +\A_{\bar p}d\bar p$
and we set, because of the glueing prescriptions for the parallel and transverse components,
$\A_{\bar\z}d\bar\z=A_{\bar\z}d\bar\z-A_{\bar p}\Gamma_{\bar\z}\bar p d\bar\z$
and $\A_{\bar p}=A_{\bar p}$, where 
$A=A_{\bar\z}d\bar\z\in T^{(0,1)}\left(S\right)$
is an anti-holomorphic 1-form on $S$, $A_{\bar p}\in\Gamma(\bar K_S^{-1})$
a section of the complex conjugate of the anti-canonical line bundle and $\Gamma_{\bar\z}d\z$
is the $(0,1)$ component of a reference connection on $\bar K_S$.

The reduction prescription is that the matrix valued dynamical fields 
$(A_{\bar\z},A_{\bar p})$ are independent of the coordinate $p$ along the fibre ${\mathbb C}$.
We now introduce a reference section $K\in\Gamma(K_S\otimes \bar K_S)$ and 
define $\phi^{(2,0)}=K A_{\bar p}\in\Gamma(K_S)$. 
With this reduction prescription and after fixing the reference connection to be $\Gamma=K^{-1}\bar\partial K$,
we get
\begin{equation}
L={\cal L}_{red}=\Omega K^{-1}\wedge Tr
\left( \frac{1}{2} \left\{
\phi^{(2,0)} \bar\partial A + A \bar\partial \phi^{(2,0)} + 2 A^2 \phi^{(2,0)}
\right\} \right) \de \bar p
\label{aredlag'}\end{equation}

To reduce to a 4-form, we saturate the reduced Lagrangian with 
$K\partial_p\wedge\partial_{\bar p}$
so that the reduced functional becomes just
\begin{equation}\label{aredlag''}
S_{red}=\frac{1}{g_s}\int_S \frac{1}{2} \Tr
\left\{ \phi^{(2,0)} \bar\partial A + A \bar\partial \phi^{(2,0)} + 2 A^2 \phi^{(2,0)}
\right\}
= \frac{1}{g_s} \int_S \Tr \left( \phi^{(2,0)} F^{(0,2)} \right)
\end{equation}
which is the form used in \cite{Diaconescu:2005jw}.

\section{The nonlinear case}

Let us now consider the reduction to the brane of the open string field theory
action on a Calabi-Yau deformation $X$ of the vector bundle $V \to \Sigma$ 
with $\Sigma \simeq {\mathbb P}^1$ a rational curve
and $V \simeq {\cal O}(n)\oplus{\cal O}(-n-2)$.
It turns out that this case
is not crucially different from the linear case.
In fact, let us note that we can relate the nonlinear and 
the linear complex structure by a singular change of coordinates. 
For the case at hand, it is enough to do it along 
the fibres above the south pole patch, namely
\begin{equation}\label{singular}
  w^i_N=\omega^i_N,\quad {\rm and} \quad w^i_S=\omega^i_S+\Psi^i\left(z_S,\omega_S\right) .
\end{equation}
In the singular coordinates $(z,w^i)$ the patching rule is the original
linear one. Therefore, Eq. (\ref{singular}) defines naturally 
the transformation rule for generic sections in the deformed complex structure
from the singular to the non singular coordinate system.
Moreover, as indicated above, the Calabi-Yau condition in the new complex structure is solved by a
potential function $X=X(z_S,\omega_S)$ such that
\begin{equation}\label{varvar}
\epsilon_{ij}w^i \de w^j = \epsilon_{ij}\omega^i \de \omega^j - \de \chi,
\end{equation}
where $w^i=\omega^i+\Psi^i(z_S,\omega)$.

That is because we can proceed by reducing the holomorphic Chern--Simons
theory in the singular coordinates (\ref{singular}) 
following the prescription of the linear undeformed case 
and then implement the variation of the complex structure
by passing to the non singular variables by field redefinition.

Let us start with the reduction in the abelian $U(1)$ case.
Then the cubic term in the Lagrangian is absent
and the reduction is almost straightforward. In the singular coordinates 
we obtain that
\begin{equation}\label{pbsing}
{\cal L}_{red}=\um\epsilon_{ij}\varphi^i\partial_{\bar z}\varphi^j 
\end{equation}
in both the north and south charts. 
The coordinate change for the fields $\varphi^i$ in terms of the ones
corresponding to the deformed complex structure is induced by
(\ref{singular}). 
Let us recall that the functions $\Psi^i$ defining the deformation
are built out from the potential $\chi$ as in eq. (\ref{varvar}).
This expresses exactly our Lagrangian terms (patch by patch) 
\begin{equation}
\epsilon_{ij} \varphi^i \partial_{\bar z} \varphi^j =
\epsilon_{ij}\phi^i \partial_{\bar z} \phi^j
- \partial_{\bar z} \chi
\label{fields}\end{equation}
where $\chi_N=0$ and $\chi_S$ is an arbitrary analytic function of the $\phi$'s
in ${\mathbb C}^2$ and of $z$ in $\mathbb{C}^*$ ($\varphi^i$ are akin to
the coordinate singular coordinate $w^i$ of the previous subsection, 
while $\phi$ stem from $\omega^i$). 

The above potential term $\chi$ gives the deformation of the action due to the
deformation of the complex structure. Specifically, we have that 
\begin{equation}\label{abred}
S_{red}=\frac{1}{g_s}
\left[
\int_{U_S} \gamma_S  ({\cal L}_{red})_S + \int_{U_N} \gamma_N  ({\cal L}_{red})_N
\right]
\end{equation}
where we explicitly indicated the resolution of the unity on the sphere 
$1=\gamma_S+\gamma_N$. For simplicity we choose the $\gamma$'s to be simply step 
functions on the two hemispheres.
Substituting (\ref{pbsing}) and (\ref{fields}) in (\ref{abred}) we then obtain
\begin{equation}
S_{red}=\frac{1}{2\,g_s}
\left[\int_{\mathbb{P}^1} \epsilon_{ij}\phi^i \partial_{\bar z} \phi^j dz d\bar z
- \int_{D} \partial_{\bar z} \chi (z,\phi) \de z \de \bar{z} \right]
\end{equation}
where $D$ is the unit disk (south hemisphere).
The disk integral can be reduced by the Stokes theorem, leaving finally
\begin{equation}
S_{red}=\frac{1}{2\,g_s}
\left[\int_{\mathbb{P}^1} \epsilon_{ij}\phi^i \partial_{\bar z} \phi^j \de z \de \bar z
+ \oint \frac{\de z}{2 \pi \ii} \chi (z,\phi) \right]
\end{equation}
where $\oint$ is a contour integral along the equator.
Therefore, we see that the reduced theory gives a $\beta$-$\gamma$ system 
on the two hemispheres with a junction interaction along the equator.

The non-abelian case is a bit more complicated than the abelian one 
because of the tensoring with the (trivial) gauge bundle. 
This promotes the vector bundle sections
to matrices and therefore unambiguously defining  
the potential function $\chi$ in the general case is not immediate.
In the following we show where the difficulty arises and which 
further constraint to the deformation of the complex structure is needed
in order to suitably deal with the non--Abelian case.

To see this let us perform the reduction
on ${\mathbb P}^1$ of the non Abelian case (\ref{hCS2}), as we did in the Abelian
case. Let us work in the singular coordinates and obtain again the pullback
Lagrangian we got in the linear case. Now, in order to pass to the non singular 
coordinates we have to promote to a matrix equation the change of variables
(\ref{fields}). This can be done by specifying a prescription for matrix 
ordering. Suppose we choose a specific ordering and denote it by $\P$.
Then our change of variable is 
\begin{equation}\label{matfields}
  \Tr \left[ \epsilon_{ij} \varphi^i \partial_{\bar{z}} \varphi^j \right] =
  \Tr \left[ \epsilon_{ij} \phi^i \partial_{\bar z} \phi^j \right] - \partial_{\bar z} \Tr \chi^\P
\end{equation}
while the cubic term gives
\begin{equation}
\Tr \left[ A_{\bar z} \epsilon_{ij} (\phi^i +\Psi^{i\P})(\phi^j + \Psi^{j\P})\right]
\label{cubicte}\end{equation}
It appears immediately that our result is complicated and seems to depend quite
non-trivially on the matrix ordering prescription. Otherwise, it is well defined.

In order to avoid matrix ordering prescriptions, 
henceforth we restrict to the case in which 
$\chi(z,\omega)$ does not depend, say, on $\omega_2$ and we proceed further. 
This is the Laufer rational case, that we shall further discuss
in chapter \ref{ch:deformations} from a geometric viewpoint.
In this case the deformation formulas simplify considerably.
Eq.(\ref{varvar}) is solved by $\Psi^1=0$ and $\Psi^2$ is determined by the 
potential by
\begin{equation}
\partial_{\omega^1}\left(\frac{\Psi^2}{\omega^1}\right)=
-\frac{\partial_{\omega^1} \chi }{(\omega^1)^2}
\label{geomat}\end{equation}
This condition can be written also as
$$2\Psi^2=\partial_{\omega^1}\left[\omega^1\Psi^2+ \chi \right].$$

As far as the reduction is concerned, the equation (\ref{matfields}) is unchanged,  
since we do not need any prescription $\P$; while eq.(\ref{cubicte}) 
simplify to
\begin{equation}\label{cubicte'}
Tr \left[A_{\bar z} \epsilon_{ij} (\phi^i +\Psi^{i})(\phi^j +\Psi^{j})\right] =
Tr \left(A_{\bar z}\left[\phi^1,\phi^2\right]\right)
\end{equation}
(where we used $[\phi^1,\Psi^2]=0$)
which, as in the linear case, is the contribution needed to complete the 
covariant derivative. 
The last operation to obtain our final result is an integration by part in the 
derivative term. As
$$\epsilon_{ij}\phi^i\partial_{\bar z}\phi^j=-2\phi^2\partial_{\bar z}\phi^1 
+\partial_{\bar z}\left(\phi^1\phi^2\right)$$ in both the north and south 
charts, from the last term we get an additional contribution to the equator 
contour integral, that is
$\frac{1}{g_s}\um\oint Tr\phi^1\Psi^2$. 
Adding it to the previously found 
term we get $\frac{1}{g_s}\um\oint \Tr( \chi +\phi^1\Psi^2)$.
This, by (\ref{geomat}) can be written just as $\frac{1}{g_s} \oint Tr B$, 
where $\partial_\omega^1 B=\Psi^2$.

Therefore, summarizing, we find that in the non Abelian case on the Riemann 
sphere  we are able to treat the deformations of the type
\begin{equation}\label{varfin}
\omega_N^1=z_S^{-n}\omega_S^1 ,
\quad{\rm and}\quad
\omega_N^2=z_S^{2+n}\left[\omega_S^2+
\partial_{\omega^1}B\left(z_S,\omega_S^1\right)\right]
\end{equation}
which corresponds to the choice $n_1=-n$. This geometry has been introduced
(in the matrix planar limit) by \cite{Ferrari:2003vp}.
These geometries are CY for any potential $B$ analytic in 
${\mathbb C}^\times\times{\mathbb C}$. The relevant reduced theory action 
is given by
\begin{equation}\label{SLauferred}
S_{red}= \frac{1}{g_s} \int_{P^1} -\Tr(\phi^2 D_{\bar z} \phi^1) +
         \frac{1}{g_s} \oint \frac{\de z}{2 \pi \ii} \Tr B(z,\phi^1) .
\end{equation}

\section{Laufer curve and matrix models}

We now show that after gauge fixing, 
the deformed $\beta$-$\gamma$ system on the Laufer rational curve 
reduces to a matrix model.

\subsection{Gauge fixing}
In order to calculate the partition function of the theory,
we now discuss the gauge fixing of the theory.
The following discussion is a refinement of the derivation given in 
\cite{Dijkgraaf:2002fc}. 
Our starting action is (\ref{SLauferred})
and we follow the standard BRST quantization (see for example \cite{Henneaux:1992ig}).

The BRST transformations in the minimal sector is 
\begin{eqnarray}
s A_{\bar z}=-(Dc)_{\bar z}, \quad
s \phi^i=[c,\phi^i], \quad
sc=\frac{1}{2}[c,c]
\end{eqnarray}
while we add a further non minimal sector to implement the gauge fixing with
$$
s\bar c=b,\quad sb=0.
$$
The gauge fixed action is obtained by adding to $S$ a gauge fixing term
$$
S_{gf}=S+s\Psi,\quad {\rm where}\quad
\Psi=\frac{1}{g_s}\int_{\mathbb{P}^1}{\Tr} \bar c \partial_z A_{\bar z}
$$
which implements a holomorphic version of the Lorentz gauge.
Actually we have
$$s\Psi=\frac{1}{g_s}\int_{P^1}{\rm Tr}\left[b\partial_z A_{\bar z}- 
\partial_z\bar c(Dc)_{\bar z}\right] .$$
The partition function is then the functional integral
$$
Z_B=\int{\cal D}\left[\phi^i,A_{\bar z},c,\bar c,b\right]
e^{-S_{gf}} .
$$
The calculation can proceed as follows.
Let us first integrate along the gauge connection $A_{\bar z}$ which 
enters linearly the gauge fixed action
and find
$$
Z_B=\int{\cal D}\left[\phi^i,c,\bar c,b\right]
e^{-\frac{1}{g_s}\left[-\int_{P^1}{\rm Tr}\phi^2\partial_{\bar z}\phi^1
- \partial_z\bar c\partial_{\bar z}c
+\oint{\rm Tr}B(z,\phi^1)\right]}
\delta \left\{
\partial_z b + [\partial_z\bar c, c] +[\phi^1,\phi^2]
\right\}
$$
Now we integrate along the field $b$. By solving the constraint we obtain
$$
Z_B=\int{\cal D}\left[\phi^i,c,\bar c\right]
e^{-\frac{1}{g_s}\left[-\int_{P^1}{\rm Tr}\phi^2\partial_{\bar z}\phi^1
- \partial_z\bar c\partial_{\bar z}c
+\oint{\rm Tr}B(z,\phi^1)\right]
}\frac{1}{{\rm det}' \partial_z}
$$
where ${\rm det}'$ is the relevant functional determinant with the exclusion of
the zero modes. Then we integrate along the $(c,\bar c)$ ghosts and get
$$
Z_B=\int{\cal D}\left[\phi^i\right]
e^{-\frac{1}{g_s}\left[-\int_{P^1}{\rm Tr}\phi^2\partial_{\bar z}\phi^1
+\oint{\rm Tr}B(z,\phi^1)\right]
}\frac{
{\rm det}' \partial_z\partial_{\bar z}
}{{\rm det}' \partial_z}
$$
Finally, since the geometric potential $B$ does not depend on $\phi^2$, 
we can also integrate along this 
variable and obtain
$$
Z_B=\int{\cal D}\left[\phi^1\right]
e^{-\frac{1}{g_s}\left[\oint{\rm Tr}B(z,\phi^1)\right]}
\delta(\partial_{\bar z}\phi^1)
\frac{
{\rm det}' \partial_z\partial_{\bar z}
}{{\rm det}' \partial_z} .
$$
The delta function constrains the field $\phi^1$ to span the 
$\partial_{\bar z}$-zero modes and once it is solved it 
produces a further $\left({\rm det}' \partial_{\bar z}\right)^{-1}$ 
multiplicative term that cancel the other determinants.
Therefore, all in all we get 
$$
Z_B=\int_{{\rm Ker}\partial_{\bar z}} d\phi^1
e^{-\frac{1}{g_s}\left[\oint{\rm Tr}B(z,\phi^1)\right]}.
$$
Lastly we can expand $\phi^1=\sum_{i=0}^n X_i\xi_i$ along the basis 
$\xi_i(z)\sim z^i$
of ${\rm Ker}\partial_{\bar z}$ with $N\times N$ matrix coefficients $X_i$.
Finally we find the multi-matrix integral
\a
Z_B=\int \prod_{i=0}^n d X_i
e^{-\frac{1}{g_s}{W}\left(X_0,\dots,X_n\right)}\label{functint2}
\b
where we defined
\a
{W}\left(X_0,\dots,X_n\right)
=\oint{\rm Tr}B(z,\sum_i X_i z^i)\label{B}
\b
This is the result of our gauge fixing procedure which confirms \cite{Ferrari:2003vp}.

The case in \cite{Dijkgraaf:2002fc} is reproduced for $n=0$. Then, the only non 
trivial complex structure deformation in (\ref{varfin}) is with 
$B=\frac{1}{z}W(\omega_1)$ (since any other dependence in $z$ can be 
re-absorbed by analytic reparametrizations)
and hence we get the one matrix model with potential $W$.

The above formula can be also inferred by just generalizing 
a CFT argument in \cite{Dijkgraaf:2002fc} to the geometry (\ref{varfin}).
To this end let us consider again the two dimensional theory defined by the 
action
\begin{eqnarray}
S = \frac{1}{g_s} \int_{{\mathbb P}^1} \Tr \left( \phi_2 D_{\bar z} \phi_1 
\right)
\end{eqnarray}
where $D_{\bar z} = \partial_{\bar{z}}+[A_{\bar{z}}, \cdot ]$. This is a 
gauged chiral conformal field theory: 
a gauged $b$--$c$ ($\beta$--$\gamma$) system in which $\phi_1$ and $\phi_2$ are 
conformal fields of dimensions $-n/2$ and $1+n/2$ respectively. 
For any $n$, the fields $\phi_1$ and $\phi_2$ are canonically conjugated and 
on the plane they satisfy
the usual OPE \begin{eqnarray}
   \phi_1 (z) \phi_2 (w) \sim \frac{g_s}{z-w}
\end{eqnarray}
In Hamiltonian formalism, that is in the radial quantization of the CFT, the 
partition function is given as
\begin{eqnarray}
Z = \left< \mathrm{out} | \mathrm{in} \right>.
\end{eqnarray}
The deformed transformation
\begin{eqnarray}
\phi_2' &=& z^{n+2} \left( \phi_2 + \partial_{\phi_1} B(z,\phi_1) \right)
\end{eqnarray}
is given on the cylinder $z=e^{w}$ as
\begin{eqnarray}
\left( \phi_2'\right)_{cyl} &=& \left(\phi_2\right)_{cyl}
                               + \frac{\partial B(z,\phi_1) }{\partial 
{\left(\phi_1\right)_{cyl}}}
\end{eqnarray}
and is implemented by the operator
\begin{eqnarray}
U = \exp \left(\Tr \oint \frac{\de z}{2 \ii \pi} B(z,\phi_1)  \right)
\end{eqnarray}
Therefore the new partition function is
\begin{eqnarray}
Z = \left< \mathrm{out} |U| \mathrm{in} \right>.
\end{eqnarray}
which is our result.

We remark that this is an {\it a posteriori} argument, it is a consistency check
but does not explain the dynamical origin of the matrix model from the string 
theory describing the brane dynamics.

\subsection{Engineering matrix models}

Once the link between D-brane configurations and multi--matrix models is 
established, the next natural question to ask is which kind of matrix 
models we get in this way. In this section we single out what is the most 
general type of multi--matrix model
we can engineer by deforming D-branes on 2--cycles in the above way and 
we produce some examples. 

The geometric potential $B(z,\omega)$ is a general holomorphic function on 
$\mathbb{C}^* \times \mathbb{C}$ 
but the terms actually contributing to a change in the complex structure 
and giving a non zero matrix potential
are of the form
\begin{eqnarray}\label{eq:geometricpotential}
B(z, \omega) = \sum_{d=1}^{\infty}  \sum_{k=0}^{d \cdot n} t^{(k)}_{d} 
z^{-k-1} \omega^d
\end{eqnarray}
where $t^{(k)}_{d}$ are the \emph{times} of the potential and $\omega$ is to 
be identified with the coordinate $\omega_1$ of the previous section.
It can be easily proven that other terms in the expansion can be re-absorbed 
by an analytic change of coordinates in the geometry and they 
do not contribute to the matrix potential.

The \emph{degree} of the potential $B$ is the maximum $d$ such that 
$ t^{(k)}_{d}$ is non-zero for some $k$ in (\ref{eq:geometricpotential}) 
and corresponds to the degree of the matrix potential, obtained as
\begin{eqnarray}\label{eq:matrixpotential}
W(X_0, \dots, X_n) = {\oint} \frac{\de z}{2 \ii \pi} 
B \left( z, \sum_{j=0}^{n} X_i z^i \right).
\end{eqnarray} 
Since this operation is linear, 
from (\ref{eq:geometricpotential}) and (\ref{eq:matrixpotential}) one gets 
a matrix potential of the form
\begin{eqnarray}\label{eq:matrixpotential2}
W(X_0, \dots, X_n) =   \sum_{d=0}^{\infty}  \sum_{k=0}^{d \cdot n} 
t^{(k)}_{d} W^{(k)}_{d}(X_0, \dots, X_n)
\end{eqnarray} 
where each term
\begin{eqnarray}
 W^{(k)}_{d}(X_0, \dots, X_n) = \sum_{i_1, \dots, i_d=0 \atop i_1+ \dots + 
 i_d=k }^n X_{i_1} \dots X_{i_d}
\end{eqnarray}
corresponds to $ B^{(k)}_{d} (z, \omega) = z^{-k-1} \omega^d$ for 
$0 \leq d \leq + \infty$ and $0 \leq k \leq{d \cdot n}$. Note that these are 
directly obtained in completely symmetric ordered form with respect to 
the indices $i_1, \dots i_d$ labeling the different matrix variables. 
In the following we will sometimes write simply the polynomial $W$ for 
c-number variables $W(x_0, \dots, x_n)$, understanding 
the total symmetrization when matrices are plugged in.

As it was already anticipated in the previous section, the {one matrix 
models} corresponds to $n=0$ and therefore $B=\frac{1}{z}W(\omega)$, \cite{Dijkgraaf:2002fc}. 

{\it Two matrix models} are obtained by considering the case $n=1$.
Some of them have been derived in \cite{Ferrari:2003vp}.
In this case, it is possible to engineer a general function for two commuting variables. In fact
\begin{eqnarray}
B(z, \omega) = z^{-k-1} \omega^{k+j} & \rightarrow & W(x)= 
{k+j \choose k} x_0^k x_1^j .
\end{eqnarray} 
and the matrix potential reads
\begin{eqnarray}
W(x_0,x_1)=
\sum_{d=1}^\infty
\sum_{k=0}^d
t_d^{(k)}{d \choose k}
x_1^k x_0^{d-k}
\end{eqnarray} 
which is, upon varying the possible couplings, a generic analytic potential 
in the two variables $x_0$ and $x_1$.
The only constraint is the matrix ordering which is always the symmetric one.
In particular, it is easy to engineer a two matrix model with bilinear 
coupling.  This is achieved by choosing, for $n=1$,
the geometric potential to be
\begin{eqnarray}
B(z, \omega) = \frac{1}{z} \left[V(\omega)+U\left(\frac{\omega}{z}\right)\right]
+\frac{c}{2z^2}\omega^2\label{2mpot}
\end{eqnarray} 
which generates the matrix potential
\begin{eqnarray}
W(x_0,x_1) = V(x_0)+U(x_1)+cx_0x_1 \label{2mpot'}
\end{eqnarray} 

In general, the {\it multi--matrix models} one can engineer are not   
of arbitrary form. Actually, on top of the fact that we can generate only 
matrix potentials with symmetric ordering, there are also constraints between 
possibly different couplings. This can be inferred from the fact that a 
polynomial function in $n+1$ variables of degree $d$ is 
specified by many more coefficients than the ones we have at our disposal. 
(As an example, if $n=3$ and $d=3$ we would need $10$ coefficients, 
while we have only $4$ at our disposal.)

To end this section, we remark that some deformations can connect 
cases with different values of $n$. The geometric equivalence of seemingly 
different complex structures becomes in fact explicit at the matrix model level.
As an example, let us consider
the case $n=2$ and a geometric potential of the form
\begin{eqnarray}
B(z, \omega) = -\frac{1}{2} z^{-4} \omega^2 + z^{-3} F(\omega).\0
\end{eqnarray} 
Out of this, one obtains
\begin{eqnarray}
W(x_0, x_1, x_2) = \left( F'(x_0) - x_1\right) x_2 + 
\frac{1}{2}\left( F''(x_0)x_1^2 \right)\0
\end{eqnarray} 
After integration of $x_2$, which appears linearly, this theory is equivalent 
to a one-matrix model with potential
\begin{eqnarray}
V(x_0) = \frac{1}{2}F''{F'}^2(x_0),\0
\end{eqnarray} 
which is equivalent to $n=0$ and $B(z,\omega)=\frac{1}{z}V(\omega)$.
In fact, the geometry with $n=2$ and $B=-\frac{1}{2} z^{-4} \omega^2 $ is 
equal, upon diagonalization, to the geometry $n=0$ and $B=0$.

\section{Flavour fields}

In the previous sections we analyzed a topological sector of
type IIB superstring theory background with space filling D5-branes 
wrapped around two-cycles of a noncompact Calabi--Yau threefold.
We also showed that in some cases the theory reduces to a matrix
model whose potential is related to deformations of complex structures of the
Calabi--Yau threefold. 

In this section we extend this analysis by including flavour fields,  
that is, we engineer $N=1$ supersymmetric gauge theories containing fields in the fundamental 
representation of the gauge group. 
More precisely we give a particular reduction prescription of
the topological open string field theory on branes in a IIB background 
(for the relation between topological
strings and superstrings, see the appendix \ref{appendix-typeII})
where $X$ is a non--compact CY given by 
${\cal O}(n)\oplus{\cal O}(-2-n)$ on $\PP^1$ with a singular point
at which an extra fibre sits.
We wrap $N$ space--filling D5-branes on 
$\PP^1$ and complete the configuration with $M$ `effective' D3-branes stuck at 
the singular point. While the D5--brane sector gives rise, as in the smooth case, to the
superpotential (and, from a geometrical point of view,
describes deformations of the smooth CY complex structure),
the effective D3-brane sector 
gives rise to a novel part of the spectrum, the corresponding superpotential data being
related to {\it linear} deformations of the CY complex structure.

We calculate the partition function for the above models 
using this reduction prescription and find that it 
reduces to multi--matrix models with flavour \cite{Argurio:2002xv,McGreevy:2002yg,Bena:2002kw}.
These are the matrix models whose resolvents have been shown to satisfy the 
generalized Konishi anomaly equations with flavour \cite{Seiberg:2002jq}.
In the $n=0$ case, the quantum superpotential in the ${\cal N}=1$ $U(N_c)$ 
gauge theory with one adjoint and $N_f$ fundamentals is obtained.
The $n=1$ case is studied in detail, in general the flavour can be integrated out and one obtains a matrix model 
with a polynomial plus a logarithmic potential. 
 
In the first section we propose the reduction of the holomorphic Chern-Simons 
to a two dimensional theory over $\PP^1$.
In following section we show that the calculation of the relevant partition function
reduces to multimatrix integrals with flavour matrices, and produce some explicit
examples. 

\subsection{The linear case}
Let us consider
the non compact CY geometries given by the total space of the vector bundle
${\cal N}={\cal O}(n)\oplus{\cal O}(-2-n)$ on $\PP^1$ augmented by an extra fibre
at a singular point of the $\PP^1$ and let us denote this space by $CY_n$.
For $n=0$ this space is the partial resolution of an $A_2$ singularity, as it 
is described in detail in the Appendix. The other cases with $n>0$ are a 
generalization thereof.

We consider type IIB theory on $R^{1,3}\times CY_n$ with 
$N$ D5-branes along $R^{1,3}\times \PP^1$ 
and $M$ 3-branes along $R^{1,3}$, the latter being stuck at a singular point 
in $\PP^1$ where the extra fibre sits.
The Calabi--Yau threefold can be topologically written as 
${\cal N} \vee \mathbf{C}^2$, 
where $\vee$ is the reduced union, i.e. the disjoint union of two spaces 
with a base point of each identified.
In the case $n=0$ we interpret this 
geometric background in the following way. We start with a resolution of an
$A_2$ singularity, see Appendix, and 
wrap $N$ D5--branes on one
cycle and $M$ and the other. When we blow down the latter, the  
$M$ D5--branes wrapped around the shrunk cycle will appear
as effective D3--branes stuck at the singular point in the remaining 
$\mathbb{P}^1$: they cannot vibrate along the base, 
while they are free to oscillate along the extra fibre. 
As we said, for generic $n$ we single out a point on $\mathbb{P}^1$ and 
add an extra fibre to render it singular, but the interpretation as blow-down of
a smooth cycle is not as evident as for $n=0$.

We would like to show that, upon topological twist, the superpotential of this theory can be calculated 
by means of the second quantized topological string theory.

The latter is given by the holomorphic Chern-Simons theory \cite{Witten:1992fb} 
(see also the lectures \cite{Marino:2004eq})
\begin{equation}
S(\A_T)=\frac{1}{g_s^2}\int_{CY_n} 
\Omega\wedge Tr_{N+M}\left(\frac{1}{2} \A_T\wedge
\bar\partial \A_T + \frac{1}{3} \A_T \wedge \A_T \wedge \A_T\right)
\label{hCS3}
\end{equation}
where $\A_T\in T^{(0,1)}\left(CY_n\right)$ with 
full Chan-Paton index $N+M$. The total string field $\A_T$ can be expanded as
\begin{equation}
\A_T = \left( 
\begin{array}{ll}
                     \A & \X \\ 
                     \tilde \X & 0 
              \end{array} \right) \nonumber
\end{equation}
where $\A$ is the string field for the 5-5 sector and $(\X,\tilde \X)$ for 
the 5-3 and 3-5
open strings. The 3-3 sector is irrelevant to us (anyway, see next footnote).

The action (\ref{hCS3}) reduces to
$$
S(\A_T)=\frac{1}{g_s^2}\int_{CY_n} 
\Omega\wedge \left[Tr\left(\frac{1}{2} \A\wedge
\bar\partial \A + \frac{1}{3} \A \wedge \A \wedge \A\right)\right.
$$
\begin{equation}
\left.
+\frac{1}{2}\sum_{I=1}^M\left(\bar D_\A\tilde\X_I \wedge \X^I +\tilde\X_I \wedge \bar D_\A \X^I\right)\right] 
\label{rhCS}
\end{equation}
where gauge indices are not shown explicitly, $I=1,\dots,M$ is the flavour index and 
$D_\A$ is the covariant derivative in the (anti-)fundamental representation.

The reduction of the open string field to the D5-brane world-volume $\mathbb{P}^1$ 
is obtained via an auxiliary  invertible bilinear form on 
${\cal N}\otimes\bar{\cal N}$ which we denote by $K$ and its associated Chern connection
$\Gamma_{\bar z}=K^{-1}\partial_{\bar z}K$.
The reduction condition for the 5-5 sector is
$(D_\Gamma\A)^{\cal N}=0$, where $D_\Gamma$ is the covariant derivative w.r.t. $\Gamma$
and the ${\cal N}$ index denotes projection along the fibre directions.
The reduction conditions for the 5-3 and 3-5 sector read 
$(D_\Gamma\X)^{\cal N}=0$ and $i_{\partial_{\bar z}}\X|_{\mathbb{P}^1}=0$,
as well as 
$(D_\Gamma\tilde\X)^{\cal N}=0$ and $i_{\partial_{\bar z}}\tilde\X|_{\mathbb{P}^1}=0$
.
The last condition specifies that the D3-branes are stuck at the singular point in
$\PP^1$
and therefore their oscillations along $T\PP^1$ are inhibited.%
\footnote{Actually, the 3-3 sector would appear as the $C$ component in
$\A_T=\left(\begin{array}{ll} \A & \X \\ \tilde \X & C \end{array}\right)$.
It would modify the action by $\Delta S=\frac{1}{g_s^2}\int_{CY_n} 
\Omega\wedge \left[Tr_{M}\left(\frac{1}{2} C\wedge
\bar\partial C + \frac{1}{3} C \wedge C \wedge C\right)
+ \tilde X \wedge C\wedge X\right]$.
Upon the reduction condition for the 3-3 sector
$i_{\partial_{\bar z}}C|_{\mathbb{P}^1}=0$, we see that the last two terms
give vanishing contribution (neither $C$ nor the $X$'s have a 
$d \bar z$ component to 
complete a top-form on $CY_n$), therefore the 3-3 sector decouples.}.

The reduction of the 5-5 sector has been already discussed in the previous sections.
The reduction of the flavour sector can be 
carried out with the same technique.
Let $X_{\bar i}^I=K_{\bar i j}Q^{jI}$ and analogously
$\tilde X_{\bar iI}=K_{\bar i j}\tilde Q^j_I$, where
$\X=X_{\bar i}dw^{\bar i}$ and 
$\tilde \X=\tilde X_{\bar i}dw^{\bar i}$
solve the above reduction conditions.
The reduction of the Lagrangian density to the $\PP^1$ for the 5-3 sector
follows the same logic as for the 5-5 sector.
The proper pullback to the base is performed patch by patch 
with the help of $K$ as contraction of the 
(3,3)-form Lagrangian by the two bi-vector fields 
$k=\frac{1}{2}
\epsilon_{ij}K^{i \bar l}K^{j \bar k}
\frac{\partial}{\partial \bar w^l}
\frac{\partial}{\partial \bar w^k}$ 
and $\rho=\frac{1}{2}
\epsilon^{ij}\frac{\partial}{\partial w^i}
\frac{\partial}{\partial w^j}$.

The resulting $(1,1)$-form Lagrangian density reads
\begin{equation}
L_{fl.red.}=
\um \epsilon_{ij}\tilde Q^i_I D_{\bar z} Q^{jI}
-
\um \epsilon_{ij}D_{\bar z}\tilde Q^i_I Q^{jI}
\label{Lrf}\end{equation}
which is independent of $K$. This proves that our reduction mechanism is well defined.
The action (\ref{Lrf}) was given in \cite{Witten:2003nn} in a similar context
and is always a $\beta\gamma$-system.
Combining with the 5-5 sector, the total reduced action then reads
\begin{equation}
L_{red.}= 
\um \epsilon_{ij}\Tr \phi^i D_{\bar z}\phi^j
+
\um \epsilon_{ij}\tilde Q^i_I D_{\bar z} Q^{jI}
-
\um \epsilon_{ij}D_{\bar z}\tilde Q^i_I Q^{jI}
\label{Lr}\end{equation}

It is straightfoward to generalize the gauge fixing procedure for the 
5-5 sector, see the discussion in the previous sections, to the total system
to show that the partition function of the theory above reduces to 
matrix integrals over the $\partial_{\bar z}$ zero--modes
of the fields.

\subsection{Deformed complex structures and matrix models}

The deformation of the complex structure of the singular spaces
we are considering can be split into two operations, namely the 
deformation of the smooth part and the deformation of the extra fibre at 
the singular point.

As far as the smooth part is concerned,
the deformed complex structures to which we confine are of the form
obtained by glueing the north and south patches of the fibres above 
the sphere as
\begin{equation}
\omega_N^1=z_S^{-n}\omega_S^1 ,
\quad{\rm and}\quad
\omega_N^2=z_S^{2+n}\left[\omega_S^2+\partial_{\omega^1}B\left(z_S,
\omega_S^1\right)\right]
\label{varfin2}\end{equation}
This, as is well--known, preserves the Calabi-Yau property of the six manifold.
As widely discussed in \cite{Bonelli:2005dc} the glueing conditions (\ref{varfin2})
gets promoted to the glueing conditions for the 5-5 sector, that is the 
chiral adjoints.

Now let us deal with the analogous deformation for the 
5-3 and 3-5 sectors. 
Let $\hat P$ be the point on $\PP^1$ where the extra fibre sits and let 
$(x^1,x^2)$ be the coordinates along the latter.
Before the complex structure deformation, the extra fibre glueing is 
given by
$$
x_N^1={\hat z_S}^{-n}x_S^1 ,
\quad{\rm and}\quad
x_N^2={\hat z_S}^{2+n}x_S^2.
$$
The complex structure deformations we confine to for this sector 
are linear and are described by the glueing conditions
\begin{equation}
x_N^1={\hat z_S}^{-n}x_S^1 ,
\quad{\rm and}\quad
x_N^2={\hat z_S}^{2+n}\left[x_S^2+M\left({\hat z}_S,\omega_S^1\right)x_S^1\right],
\label{varvec}\end{equation}
where $M$ is locally analytic on ${\bf C}\times\left(U_N\cap U_S\right)$.
This can be cast in the form
\begin{eqnarray}
M(z, \omega) = \sum_{d=1}^{\infty} \sum_{k=0}^{n d+2}  m_d^{k} z^{-k-1} 
\omega^d .
\end{eqnarray}
Notice however that only a subset of the parameters $m_d^{k}$ parameterize 
actual deformations of the complex structure, since only a part of them cannot 
be reabsorbed by a local reparametrization.

The deformed glueing condition (\ref{varvec}) is coupled to the 3-5 sector
of the topological open string field since the 3-branes only vibrate 
transversely along the extra-fibre.

Note that we are obtaining different string backgrounds on our geometry, by 
considering the smooth variety (\ref{varfin2}) 
and `attaching' to it additional fibres, eq. (\ref{varvec}).
The function $M$ then generates the variation of the complex structure of 
the CY along the singular fibre.

This deformed geometry (\ref{varvec}) can be implemented in the reduction of 
the open string field in a way much similar to the one followed for the pure 
5-5 sector in the smooth case. 
This is done by promoting (\ref{varfin2}-\ref{varvec}) to the glueing 
conditions of the reduced string field components $\phi^i$, $X^i$ and 
$\tilde X^i$ with a patch by patch singular field redefinition which reabsorbs
the deformation terms $(B,M)$.
In these singular coordinates the fields glue linearly and we can apply the 
results of the previous section, obtaining in this way the Lagrangian 
(\ref{Lr}) in the singular field coordinates.
Going back to the regular coordinates, one gets the action
$$
S_{red.}=\int_{\PP^1} 
\Tr \left[\phi^2 D_{\bar z}\phi^1
+
\tilde Q^2_I D_{\bar z} Q^{1I}
+
Q^{I2} D_{\bar z} \tilde Q^1_I\right]
$$
\begin{equation}
+\oint_{\mathcal{C}_0} \de z \left[\Tr B(\phi^1,z)+ \tilde Q_I^1 M(\phi^1,z) Q^{1I}\right].
\label{Lrd}\end{equation}

The partition function of the latter theory can be calculated 
as in \cite{Bonelli:2005gt} and one gets as a result a multi matrix--model of vector 
type, namely with additional interactions with flavours.
The fields $\phi^1$ and $(Q^1,\tilde Q^1)$ contribute only through
their $\partial_{\bar z}$
zero--modes. Under the above glueing prescription we expand
$\phi^1(z)=\sum_{i=0}^n z^i X_i$
and analogously
$Q^{1I}(z)=\sum_{i=0}^n z^i q^I_i$
and
$\tilde Q^1_I(z)=\sum_{i=0}^n z^i \tilde q_{iI}$, where $X_i$, $q^I_i$ and 
$\tilde q_{iI}$ are matrix
and vector constant coefficients.

Specifically the partition function, after the above calculations, reads
\begin{equation}
Z_n=\int \prod_{i=0}^n dX_i dq_i d\tilde q_i\,
{\rm e}^{\frac{1}{g_s^2}\left(\Tr {\cal W}(X) + \tilde q^I_i {\cal M}(X)_{ij} 
q_{Ij}\right)}
\label{vmmm}\end{equation}
where 
$${\cal W}(X)=\oint dz\, B\left(\sum_{i=0}^n z^i X_i,z\right)$$
is the 5-5 contribution already obtained in \cite{Bonelli:2005dc}
and
$$
{\cal M}(X)_{ij}=\oint dz\, z^{i+j}M\left(\sum_{k=0}^n z^k X_k,z\right)
$$
represents the 5-3/3-5 coupling. 

We are therefore finding in the general case a multi matrix model with flavour symmetry.
As explained in \cite{Argurio:2003ym},
these matrix models 
are related to the quantum superpotential of a ${\cal N}=1$ SYM  
with
$(n+1)M$ chiral multiplets in the fundamental/anti-fundamental and $n+1$ in 
the adjoint representation.

Let us specify a couple of examples which turn out to be interesting.

\paragraph{}

In particular, if $n=0$ and the CY manifold is the total space of 
$\left[{\cal O}(-2)_{\PP^1}\vee {\bf C}\right]\times {\bf C} $ we have a single set of 
constant zero--modes.
Choosing 
$$
B(\omega^1,z)=\frac{1}{z}{\cal W}(\omega^1)
\quad {\rm and} \quad
M(\omega^1,z)=\frac{1}{z}{\cal M}(\omega^1)
$$
we find
\begin{equation}
Z_0=\int dX dq d\tilde q\,
{\rm e}^{\frac{1}{g_s^2}\left(\Tr {\cal W}(X) + 
\tilde q^I {\cal M}(X) q_{I}\right)}.\label{Z0}
\end{equation}

This partition function is then an extended matrix model with vector
entries of the same type as the ones first considered in 
\cite{Argurio:2002xv}, which gives rise, in the
large $N$ expansion, to the quantum superpotential for ${\cal N}=1$ with 
$M$ flavour.
Actually a deeper analysis of these models, started soon after \cite{Bena:2002kw},
culminated in \cite{Seiberg:2002jq}, where it was shown that the resolvent of the 
above enriched matrix model (\ref{Z0}) solves the generalized Konishi 
anomaly equations of the corresponding four dimensional gauge theory.
It is natural therefore to conjecture that the same is true for the 
other matrix models (\ref{vmmm}) that we have just obtained above.

In particular, if ${\cal M}(\omega^1)=\omega^1-m$ and 
${\cal W}'(\omega^1)=(\omega^1)^{N}+\dots$, 
the correct SW curve
$$
y^2=[{\cal W}'(x)]^2 +\Lambda^{2N-N_f}(x-m)^{N_f}
$$
is recovered \cite{McGreevy:2002yg}.
The above SW curve should be related to the deformed partially 
resolved geometry we are considering.

\paragraph{} 

As a further example, let us discuss the result we obtain in the $n=1$ case.
Let us denote by $\Phi_0$ and $\Phi_1$ the two adjoint chiral superfields,
then the formulas for the superpotential and for the mass term read
\begin{eqnarray}
\mathcal{W}(\Phi_0, \Phi_1) &=& \sum_{d=1}^{\infty} \sum_{k=0}^{d} 
                           t^{k}_d 
                           \sum_{i_1, \dots i_d = 0,1 \atop i_1 + \dots + 
			   i_d =k} \Phi_{i_1} \dots \Phi_{i_d} \nonumber \\ 
\mathcal{M}(\Phi_0, \Phi_1) &=& \sum_{d=1}^{\infty} \sum_{k=0}^{d} 
          \left(\begin{array}{ll}
             m_d^{k} &  m_d^{k+1}  \\
             m_d^{k+1} & m_d^{k+2} 
          \end{array} \right)
                       \sum_{i_1, \dots i_d = 0,1 \atop i_1 + \dots + i_d =k} 
		       \Phi_{i_1} \dots \Phi_{i_d} \ . 
\end{eqnarray}

Note that it is possible to produce a superpotential and a 
mass term of 
the form (with $X \equiv \Phi_0$, $Y \equiv \Phi_1$)
\begin{eqnarray}
\mathcal{W}(X,Y) &=& V(X) + t' Y^2 + cXY \nonumber \\ 
\mathcal{M}(X,Y) &=& 
   \left(\begin{array}{ll}
            \mathcal{M}_1(X) & 0 \\
             0 & 0 
          \end{array} \right)
\end{eqnarray}
by considering the following geometric deformation terms
\begin{eqnarray}
B(z, \omega) &=& \frac{1}{z} V(\omega) + \frac{t'}{z^3} \omega^2 + 
\frac{c}{2 z^2} \omega^2 \nonumber \\
M(z, \omega) &=& \frac{1}{z} \mathcal{M}_1(\omega)  \ .
\end{eqnarray}


\chapter[Versal deformations for the Laufer curve]{Versal deformations \\ for the Laufer curve}
\label{ch:deformations}

In this chapter we consider 
the probem of finding 
the tree-level superpotential for a B-brane on a curve 
in a Calabi--Yau threefold as a problem in deformation theory, 
see \cite{Bershadsky:1995qy}. 
This analysis will be very much in the spirit of 
\cite{Katz:2000ab} and \cite{Clemens:2002}. 

Let us briefly recall the properties of the deformation of a curve inside a variety with trivial canonical bundle. 
The infinitesimal deformation theory of a curve $\Sigma$ inside a threefold $X$ 
is described by a holomorphic map
\begin{eqnarray}
K: U \to H^1(\Sigma, N)
\end{eqnarray}
where $U \subset H^0(\Sigma, N)$ is a neighborhood of $0$ and $N$ is the normal bundle to the curve. 
Now, if $X$ is a Calabi--Yau threefold, then by Serre duality $H^1(\Sigma, N) \simeq H^0(\Sigma, N)^*$.
This is an example of a symmetric deformation theory, \cite{behrend-2005}.
Since $K$ is a map from a vector space to its dual,
it can be seen as a 1-form on the vector space $H^0(\Sigma, N)$.
In this context, the \emph{superpotential} $W$ is a holomorphic map
\begin{eqnarray}
W: U \to \mathbb{C}
\end{eqnarray}
such that $K = \de W$. 
Notice also that, since vector spaces are cohomologically trivial, the existence of a superpotential
is equivalent to the condition $\de K = 0$. Choosing a basis of sections and coordinates
$x:=(x^1, \dots, x^h)$ on $H^0(\Sigma, N)$ and given the superpotential $W=W(x^1, \dots x^h)$,
one can write $K$ as
\begin{eqnarray}
  K = \sum_{i=1}^{h} K_i (x) \de x^{i} = \sum_{i=1}^{h} \frac{\partial W (x)}{\partial x^{i}} \de x^{i} .
\end{eqnarray}

In particular, we will consider a class of Calabi--Yau threefolds 
and embeddings, the \emph{Laufer curves}, 
that were described in chapter \ref{ch:localCY}, section \ref{sec:Laufercurves}. 
The peculiarity of the Laufer case is the ``linearity'' of the deformation theory;
in this case, the infinitesimal deformations are also finite ones and 
they correspond to actual holomorphic sections of the fibre-bundle. 
Following \cite{Bruzzo:2005dp,Bruzzo:WIP}, we will propose a relation between the 
geometric potential of the Laufer curve and the superpotential of the deformation theory.
Also, in the rational case we will prove a conjecture by Ferrari \cite{Ferrari:2003vp}, connecting the Hessian of the superpotential to the normal bundle of the curve inside the threefold.

Note that the superpotential was originally computed in the $C^\infty$ category as an extension of the Abel-Jacobi mapping in \cite{Witten:1997ep}, and see also \cite{Martucci:2006ij} for a quite general derivation. 	
For the definitions and the properties of (some of) the geometrical objects used in this chapter, see the appendix \ref{appendix-geometry}, in particular the section \ref{appendix-families}.

\section{Superpotential and geometric potential}

In this section we propose \cite{Bruzzo:WIP} a construction of the superpotential $W$ from the Laufer geometric potential $B$.
Recall from chapter \ref{ch:localCY}, that the transition functions for a Laufer curve are
\begin{eqnarray}\label{eq:lauferfibration-recall}
\left\{\begin{array}{rcl}
   z_{\alpha}  &=& f_{\alpha \beta} (z_\beta)  \\ 
   \omega^1_{\alpha} &=& \phi_{\alpha\beta} (z_\beta) \,  \omega^1_{\beta}  \\
   \omega^2_{\alpha} &=& \phi'_{\alpha\beta} (z_\beta) \left( \omega^2_{\beta} + \partial_{\omega^1_{\beta}} B_{\alpha\beta}(z_\beta, \omega^1_{\beta}) \right)
\end{array}\right.
\end{eqnarray}
where $\phi$ is a line bundle on a smooth curve $\Sigma$ and $\phi' = K_\Sigma \phi^{-1}$.
$B$ is the geometric pontential and can be understood as an element of the (infinite-dimensional) space 
\begin{eqnarray}
  \bigoplus_{d=1}^\infty H^0(\Sigma, \phi^d)^*
\end{eqnarray}
where, as before, we will not address convergence issues when discussing power series.
We also assume the following conditions on the line bundle $\phi$
\begin{eqnarray}\label{eq:additionalconditions}
  H^0(\Sigma, \phi) &>& 0 \nonumber \\
  H^1(\Sigma, \phi) &=& 0 
\end{eqnarray}
and define $h:= \dim H^0(\Sigma, \phi)$ and $h^{(d)}:= \dim H^0(\Sigma, \phi^d)$.
Using these conditions and the Riemann-Roch theorem, we obtain the following relations
\begin{eqnarray}
h &=& \deg (\phi) - g + 1 \nonumber \\
h^{d} &=& d \deg (\phi) -g + 1 = d h + (d-1)(g-1) \ .
\end{eqnarray}

Remark now that, after choosing a basis in the space of the sections of $\phi$, and since $\phi'$ has no sections, the superpotential is a holomorphic function of $h$ complex variables, where $h$ is the dimension of $H^0(\Sigma, \phi)$. If we expand it in polynomials of fixed degree $d$
\begin{eqnarray}
W(x_1, \dots, x_h) = \sum_{d=1}^{\infty} w_d(x_1, \dots, x_h) ,
\end{eqnarray}
then each $w_d$ can be regarded as a multilinear functional on the sections, so an element in $H^0(\Sigma, \phi)^{*\otimes d}$. 
Note also that 
\begin{eqnarray}
   \dim \mathrm{Sym}^d H^0(\Sigma, \phi) = \left( {h+d-1 \atop d}  \right) \ .
\end{eqnarray}
Finally, notice that we also have a map (the \emph{multiplication of sections morphism})
\begin{eqnarray}
\mu : H^0(\Sigma, \phi)^{\otimes d} \to H^0(\Sigma, \phi^d) .
\end{eqnarray}

We can now propose the construction of the relevant superpotential for the Laufer curve. 

\begin{proposition}
Let $X \to \Sigma$ be a Laufer curve and assume that the condition of equation (\ref{eq:additionalconditions}) is satisfied. Then, 
\begin{enumerate}
\item 
the holomorphic sections of the fibre-bundle are in a one-to-one correspondence with the critical points of a holomorphic function $W$, the superpotential, \emph{i.e.}, with the  solutions of the equations
\begin{eqnarray}
\frac{\partial W}{\partial x_i} = 0 \ , \quad i=1, \dots, h
\end{eqnarray} 
where each degree-$d$ term $w_d$ of $W$ is obtained from the expansion of the geometric potential $B$
by applying the map
\begin{eqnarray}
\mu^* : H^0(\Sigma, \phi^d)^* \to H^0(\Sigma, \phi)^{*\otimes d}
\end{eqnarray}
dual to the multiplication of sections morphism.
\item 
the normal bundle to the curve defined by a critical point of the superpotential $W$ is determined by the Hessian of $W$ at the given critical point.
\end{enumerate}
\end{proposition}
We do not give a proof of the proposed statement, 
that generalizes the results known in the rational case, 
\cite{Katz:2000ab, Ferrari:2003vp};
hopefully it will appear in \cite{Bruzzo:WIP}. 
In the following we specialize to the rational case, and prove
a ``version'' of the second part of the previous proposition, 
first conjectured in \cite{Ferrari:2003vp}.

\section{Rational case and Ferrari's conjecture}

Let $\cC \simeq\PP^1$ be a smooth rational curve and $V \to \cC$ a rank-2 holomorphic vector bundle on $\cC$, with $\det V \simeq K_{\cC} \simeq \cO(-2)$, so that  the total space of the bundle $V$  has trivial canonical bundle. Then $V \simeq  \cO(-n-2) \oplus \cO(n)$ for some $n$. We  consider deformations of $V$ given in terms of transition functions
in the standard atlas $\mathcal{U} = \{ U_0, U_1 \}$ of $\mathbb{P}^1$ 
 as 
\begin{equation}\label{eq:fibration}
\left\{\begin{array}{rcl}   z'  &=& 1/z  \\[3pt]   \omega_1' &=& z^{-n}  \omega_1   \\[3pt]   \omega_2' &=& z^{n+2} \left( \omega_2 + \partial_{\omega_1} B(z, \omega_1) \right)\ .\end{array}\right.
\end{equation}
Note that the complex manifold $X$ defined as the total space of this fibration has again trivial canonical bundle.  The term $B(z, \omega)$ is a holomorphic function on $(U_0 \cap U_1) \times \mathbb{C}$ and is called the \emph{geometric potential}. If we expand the function $B$ in its second variable
\begin{equation}
B(z,\omega_1)=\sum_{d=1}^\infty \sigma_d(z)\,\omega_1^d
\end{equation}
each coefficient $\sigma_d$ may be regarded as a cocycle
defining an element in the group
\begin{equation}
H^1(\PP^1,\cO(-2-dn))\simeq H^0(\PP^1,\cO(nd))^\ast\,.
\end{equation}

\paragraph{}
If we consider $\cC$ as embedded in $X$ as a section, 
and consider the problem of deforming the embedding $\cC \to X$, 
the space of versal deformations can be conveniently described 
by a superpotential \cite{Katz:2000ab}. 
In the case at hand the  superpotential $W$ can be defined as
the function of $n+1$ complex variables given by
\begin{equation}\label{eq:superpotential}     
W(x_0, \dots, x_n) = \frac{1}{2 \pi \mathrm{i}} \oint_{\cC_0} B\left(z,\omega_1(z)  \right) \,\de z
\end{equation}
where $z$ and $z'$ are local coordinates on $U_0$ and $U_1$,
and the parameters $x_0, \dots, x_n$ define sections of 
 the   line bundle $\cO(n)$ by letting 
\begin{eqnarray}\label{eq:section1}
   \omega_1(z)= \sum_{i=0}^{n} x_i z^i \ , \qquad \omega_1'(z') = \sum_{i=0}^{n} x_i (z')^{n-i}\,.
\end{eqnarray} 
One should note that the superpotential $W$ can be obtained by applying
to the function $B$, regarded as an element in $H^0(\PP^1,\cO(nd))^\ast$, the dual of the
multiplication morphism
\begin{equation}
H^0(\PP^1,\cO(n))^{\otimes d} \to H^0(\PP^1,\cO(nd))
\end{equation}
(here one should regard the dual of $H^0(\PP^1,\cO(nd))$ as a space
of Laurent tails).

The key to the result  we want to prove is the relationship occuring between
the superpotential $W$ and the sections of the fibration $X\to \cC$
(cf.~\cite{Katz:2000ab, Ferrari:2003vp}).
\begin{lemma}\label{lemma:sect}The holomorphic sections of the   fibration $X\to \cC$ 
are in a one-to-one correspondence with the critical points of the superpotential, \emph{i.e.}, with the  solutions of the equations\begin{eqnarray}\frac{\partial W}{\partial x_i} = 0 \ , \quad i=0, \dots, n \ .\end{eqnarray} \end{lemma}
\begin{proof}
This  can be verified by explicit calculations \cite{Ferrari:2003vp} after representing  the sections    of $X$ as
\begin{equation}\label{eq:section2}\begin{array}{rcl}  \omega_2(z) &=& \displaystyle  -  \frac{1}{2 \ii \pi}\oint_{C_z} \frac{\partial_\omega B (u,\omega_1(u))}{u-z}  \,  \de u \\[12pt] \omega_2'(z') &=& \displaystyle  \frac{1}{2 \ii \pi} \oint_{C_{z'}}
 \frac{\partial_\omega B (1/u,\omega_1(1/u))}{u^{n+2}(u-z)}  \,\de u\end{array}\end{equation}where the contour $C_z$ (resp. $C_{z'}$)  encircles the points $0$ and $z$ (resp $z'$). So (\ref{eq:section1}) and (\ref{eq:section2}) yield a rational curve $\Sigma \subset X$ for each critical point $(x_0, \dots, x_n)$ of $W$.
\end{proof}

\paragraph{} 
Now we state and prove Ferrari's conjecture.
\begin{proposition} \label{prop:ferrari}
The normal bundle to the section $\Sigma$ of $X$ determined by a critical point $(x_0, \dots, x_n)$ of $W$ is $\cO_\Sigma(-r-1) \oplus \cO_\Sigma(r-1)$ where $r$ is the corank of the Hessian of $W$ at that point.
\end{proposition}
To calculate the normal bundle to $\Sigma$ we first need to linearize the transition functions around the given section. Defining new coordinates $\delta_i = \omega_i - \omega_i(z)$, $\delta_i' = \omega_i' - \omega_i'(z)$, we obtain
\begin{eqnarray}  
\delta_2' =  z^{n+2} \left(\delta_2 +  h(z) \delta_1 + g(z) \right)
\end{eqnarray}
where 
\begin{eqnarray}    
g(z) = \partial_\omega B(z, \omega_1(z)) \ ,  \qquad   h(z) =  \partial^2_\omega B(z, \omega_1(z)) 
\end{eqnarray}
and at a critical point of $W$ we have $g(z) = 0$ using relation (\ref{eq:derivatives}) in the next section.
Furthermore, again from (\ref{eq:derivatives}), for $h(z)$ we have
\begin{eqnarray}\label{eq:cocycleW}
  h(z) = \sum_{i \leq j =0}^{n} \partial_i \partial_j W^{(k)}_{d} z^{-(i+j)-1}
\end{eqnarray}
up to terms that can be can be reabsorbed by a holomorphic change of coordinates (see the next section).
Now we need the following. Let us consider an extension of vector bundles on $\PP^1$ of the form
\begin{eqnarray}
0 \longrightarrow \cO_{\PP^1}(-n-2) \longrightarrow \Phi   \longrightarrow \cO_{\PP^1}(n) \longrightarrow 0
\end{eqnarray}
parametrized by a cocycle $\sigma \in H^1 (\PP^1, \cO_{\PP^1}(-2n-2))$. 
With respect to the two standard charts $U_0, U_1$ and in the coordinate $z$ of $U_0$,  $\sigma$ can be written as
\begin{eqnarray}
   \sigma(z) = \sum_{k=0}^{2n} \widetilde{t}_k z^{-k-1} \ .
\end{eqnarray}
Let us define a quadratic form (quadratic superpotential)on the global sections   of  the line bundle $\cO_{\PP^1}(n)$:
\begin{eqnarray}   H(x_0, \dots, x_n) = \sum_{k=0}^{2n} \widetilde{t}_k \sum_{i,j=0 \atop i+j=k }^{n} x_i x_{j} = \sum_{i,j=0}^{n} H_{ij} x_i x_j \ .\end{eqnarray}\begin{lemma}The vector bundle $\Phi$ is $\cO_{\mathbb{P}^1}(r-1) \oplus \cO_{\PP^1}(-r-1)$, where $r$ is the corank of the quadratic form $H$.\end{lemma}\begin{proof} By Lemma \ref{lemma:sect} the sections of the bundle $\Phi$ correspond to the critical points of $H$, \emph{i.e.,} to the solutions of the linear system\begin{eqnarray}   \sum_{j=0}^{n} H_{ij} x_j = 0 \ .\end{eqnarray}The dimension of this space is $r$, the corank of $H$. The only rank two vector bundle over $\mathbb{P}^1$ with determinant $\cO_{\PP^1}(-2)$ and $r$ linearly indipendent holomorphic sections is $\cO_{\PP^1}(r-1) \oplus \cO_{\mathbb{P}^1}(-r-1)$.  \end{proof}
The proof of   Proposition \ref{prop:ferrari} is now complete: 
in fact, by (\ref{eq:cocycleW}) the quadratic form $H$ corresponds to the Hessian of the superpotential $W$ at its critical points.

\subsection*{Some formulas for the potentials}
We group here some formulas that turn out to be useful in checking the
computations involved in the results presented in this paper.
\paragraph{} 
The geometric potential (deformation term) $B(z, \omega_1)$ is holomorphic on $\mathbb{C}^* \times \mathbb{C}$ and can be cast in the form
\begin{eqnarray}\label{eq:expansionB}
 B(z, \omega) =  \sum_{d=0}^{\infty} \sum_{k=0}^{dn} t^{(k)}_d  B^{(k)}_d (z, \omega)
\end{eqnarray}
where 
\begin{eqnarray}
 B^{(k)}_d (z, \omega) = z^{-k-1} \omega^d \qquad k= 0, \dots, dn \ .
\end{eqnarray}
The terms with $k < 0$ or $k >  dn$ can be reabsorbed by a holomorphic change of coordinates. For $l := - k -1 \geq 0$, we define $\widetilde{\omega}_2 := \omega_2 + d z^{l} \omega_1^{d-1}$, and for $m := k - dn -1 \geq 0$, we define
\begin{eqnarray}
\widetilde{\omega}_2':= \omega_2' - (z')^m (\omega_1')^{d-1}\ .
\end{eqnarray}
%

The superpotential that corresponds to $B^{(k)}_d$, given by (\ref{eq:superpotential}), is
\begin{eqnarray}\label{eq:expansionW}
 W^{(k)}_{d}(x_0, \dots, x_n) = \sum_{i_1, \dots, i_d=0 \atop i_1+ \dots + i_d=k }^n 
                                x_{i_1} \dots x_{i_d} \ .
\end{eqnarray}
We can obtain simple relations for the derivatives of these polynomials:
\begin{eqnarray}
 \frac{\partial W^{(k)}_{d}}{\partial x_j}  &=& \sum_{i_1, \dots, i_d=0 \atop i_1+ \dots +  i_d=k }^n 
                                                d \left( \frac{\partial x_{i_1}}{\partial x_j} x_{i_2} \dots x_{i_d} \right) \nonumber \\
                                            &=& d \sum_{i_1, \dots, i_{d-1}=0 \atop i_1+ \dots +  i_{d-1}=k-j }^n x_{i_1} 
                                                \dots x_{i_{d-1}} 
                                            \ = \ d W^{(k-j)}_{d-1}
\end{eqnarray}
and in general we have
\begin{eqnarray}
 \frac{\partial}{\partial x_{j_1}}\dots \frac{\partial}{\partial x_{j_l}} W^{(k)}_{d} = d (d-1)\dots (d-l+1) W^{(k -j_1\dots- j_l)}_{d-l}
\end{eqnarray}

\paragraph{}
Given a section $\omega_1(z)$, we have
\begin{eqnarray}\label{eq:derivatives}
\partial_\omega B(z, \omega_1(z))  = \sum_{j=0}^n \frac{\partial W}{\partial x_j} z^{-j-1} + \mathrm{trivial\ terms} \nonumber \\
\partial^2_\omega B(z, \omega_1(z)) = \sum_{i \leq j =0}^{n} \partial_i \partial_j W z^{-(i+j)-1} + \mathrm{trivial\ terms}
\end{eqnarray}
where the ``trivial terms'' can be readsorbed by a holomorphic change of coordinates. We can obtain these results from (\ref{eq:expansionB}) and (\ref{eq:expansionW}). We have
\begin{eqnarray}
\partial_\omega B^{(k)}_d (z, \omega_1(z)) = 
       d \sum _{i_1, \dots, i_{d-1} = 0}^{n} x_{i_1} \dots  x_{i_{d-1}} z^{i_1 + \dots + i_{d-1}-k-1}
\end{eqnarray}
and the only non-trivial terms are such that $0 \leq - (i_1 + \dots + i_{d-1}-k) \leq n$.
In the same way, for the second derivatives we have
\begin{eqnarray}
     \partial^2_\omega B^{(k)}_{d}(z, \omega_1(z)) = 
           d (d-1) \sum _{i_1, \dots, i_{d-2} = 0}^{n} x_{i_1} \dots  x_{i_{d-2}} z^{i_1 + \dots + i_{d-2}-k-1}
\end{eqnarray}
The relevant terms are those with  $0 \leq - (i_1 + \dots + i_{d-1}-k) \leq 2 n$.

\chapter{Conclusions and outlook}

Many questions remain open and are left to further investigation.

In particular, it would be very interesting to study the moduli space of the local curve, 
and to find the superpotential and the open string field theory for the general local curve. 
Also the geometries leading to flavour fields would need a better understanding.

Also, as indicated above, gauge/gravity correspondence and large-$N$ duality for the local curve in higher genus is somehow the motivation of the present work. 
For this reason, let us speculate on the possible structure of this large-$N$ dual. 

Let us first remark that large-$N$ transition are related to the renormalization group flow, basically because when ``summing over the boundaries'' one is integrating out some degrees of freedom of the theory. As discussed for example by \cite{Ferrari:2003vp}, one expects an extremal transition of the kind of the conifold transition and its generalizations only in the cases in which the corresponding gauge theory is asymptotically free (and confining). The volume of the curve going to zero corresponds to the fact that at the QCD scale the gauge coupling constant goes to infinity. At lower energies the theory should be effectively described by the glueball superfields. 
On the other hand, in this construction asymptotic freedom is connected to the number of holomorphic sections of the normal bundle to the curve. 

Of course it may well happen that no geometric dual does exist for the local curve in higher genus. Assuming that it does exist, let us try to guess its form by generalizing the conifold case.
We are searching for a local Calabi--Yau threefold $X_{dual}$ with the following properties. It should have the same ``asymptotics'' of the local curve, that is, topologically on should have $\partial X_{dual} \simeq \Sigma \times S^3$. 
Whether the local curve has a non trivial 2-cycle $\Sigma$ and a trivial 3-cycle $S^3$ (the boundary of the fibre $\mathbb{C}^2$) linking one another, the dual space should have a trivial 2-cycle $\Sigma$ and a non-trivial 3-cycle $S^3$. 
Let $M_\Sigma$ be the 3-cycle such that $\partial M_\Sigma = \Sigma$, the easiest case being the one in which $M_\Sigma$ is the handlebody of $\Sigma$ without boundaries, in the same way in which $\mathbb{R}^3$ is the handlebody of $\mathbb{P}^1 \simeq S^2$ without boundaries. Could it be just $X_{dual} \simeq M_\Sigma \times S^3$? In the conifold case the dual Calabi--Yau threefold topologically turns out to be $T^*S^3 \simeq \mathbb{R}^3 \times S^3$. The topological Calabi--Yau condition $c_1 = 0$ seems to suggest the same structure also for the more general case, since the handlebody constructed from a Riemann surface is ``flat''.
We do not know whether the proposed $X_{dual}$ actually admits a complex structure with trivial canonical bundle. Also we have not to forget that the integral of the eventual holomorphic $(3,0)$-form on $S^3$ has to be equal to $t:=N g_s$.

At any rate, even in the case in which no geometric dual does exist, it could be interesting to understand this non-geometric phase of the Calabi--Yau moduli space.
Note also that geometric transitions are usually symmetric in the A-model and the B-model in the sense that the open side and the closed side of the duality are exchanged in the two models.

\paragraph{}
Very interesting new developments \cite{Kapustin:2004gv,Kapustin:2005uy,Bredthauer:2006hf} are related to $(2,2)$ twisted {$\sigma$--models on generalized complex manifolds}, see \cite{Hitchin:2004ut,Gualtieri:2003dx} for an introduction to these geometries. In this setting one wants to answer questions about the category of branes, mirror symmetry, open/closed large-$N$ dualities and the relation with the physical theory; an answer to these questions could also shed light on the same problems in the ``standard'' A and B models.

Another promising research direction is given by the embedding of both A--model and B--model in a {topological M-theory}, either given as a second-quantized theory \cite{Gerasimov:2004yx, Dijkgraaf:2004te, Nekrasov:2004vv, Pestun:2005rp} or as a topological brane first-quantized description \cite{Bonelli:2005rw}.


\appendix

\chapter{The relation with Type II backgrounds}\label{appendix-typeII}

Topological strings can be considered as the topological sector of Type II strings. In this appendix we give some details of this relation, following \cite{Neitzke:2004ni}.

\section{Closed background without fluxes: $\mathcal{N}=2$ supersymmetry}
Let us consider Type II (either IIA or IIB) theory compactified on a Calabi--Yau threefold, i.e. with target space $\mathbb{R}^{3,1} \times X$, where $X$ is a Calabi--Yau threefold. The holonomy of $X$ breaks $3/4$ of the original supersymmetry in $d=10$, leaving $8$ supercharges and an $\mathcal{N}=2$ algebra in $4$ dimensions. The massless field content in $4$ dimensions can be organized in multiplets of $\mathcal{N}=2$ supergravity, according to the following table
\begin{center}
\begin{tabular}{|l|l|l|l|}
\hline
            & vector       & hyper            & gravity \\
\hline            
IIA on $X$  & $h^{1,1}(X)$ & $h^{2,1}(X)+1$ & $1$ \\
IIB on $X$  & $h^{2,1}(X)$ & $h^{1,1}(X)+1$ & $1$ \\
\hline
\end{tabular}
\end{center}
Each vector multiplet contains a single complex scalar, that corresponds to the K\"ahler moduli of $X$ in the IIA case and to the complex structure moduli of $X$ in the IIB case. The topological string partition function at each genus computes particular F--terms in the low energy effective action for supergravity, involving the vector multiplets \cite{Antoniadis:1993ze,Bershadsky:1993cx}. These F--terms can be written in terms of the $\mathcal{N}=2$ Weyl multiplet, a chiral superfield $\mathcal{W}_{\alpha \beta}$ with lowest component the self-dual part of the ``graviphoton'' $F_{\alpha \beta}$ as 
\begin{eqnarray}
   S_g = \int \de^4x \int \de^4 \theta F_g (X^I) \left( \mathcal{W}^2 \right)^g
\end{eqnarray}
where
\begin{eqnarray}
\mathcal{W}^2 := \mathcal{W}_{\alpha \beta} \mathcal{W}_{\alpha' \beta'} 
                 \epsilon^{\alpha \alpha'} \epsilon^{\beta \beta'} 
\end{eqnarray}
and $F_g(X^I)$ is the genus $g$ topological string free energy, written as a function of the vector multiplet $X^I$, i.e. for the Type IIA, $F_g$ is the A--model free energy and for Type IIB, it is the B--model free energy. Note that each $F_g$ contributes to a different term in the effective action of supergravity, writing in components we obtain for example a term of the form
\begin{eqnarray}
   \int \de^4 x F_g(X^I) \left( R_{+}^2 F_{+}^{2g-2} \right)
\end{eqnarray}
so $F_g$ represents the gravitational correction to the amplitude for the scattering of $2g-2$ graviphotons. In particular, $F_0$ is the prepotential of the gauge theory.

\section{Branes and fluxes: $\mathcal{N}=1$ supersymmetry}

Starting with the $\mathcal{N}=2$ supersymmetric compactification of the previous section, one can further reduce supersymmetry by half in two ways. The first way is to consider open strings with branes, the second to consider closed backgrounds with Ramond-Ramond 3-form fluxes. To have Poincar\'e invariance in four dimensions, the branes have to fill the four dimensions and the fluxes have to be only in the Calabi--Yau directions. The two are connected by a geometric transition in Type II strings. In this case, the topological strings compute the effective superpotential, that determines much of the infra-red behaviour of the theory.

\subsection{The open side with branes}

If we consider $N$ D-branes on cycles of the Calabi--Yau manifold and filling the four dimensions of spacetime, we obtain an $\mathcal{N}=1$ theory in 4 dimensions with $U(N)$ gauge symmetry. The open topological strings are given by the $F_{g,h}$ free energies for maps from a genus $g$ Riemann surface with $h$ boundaries such that each boundary is mapped to one D-brane. The relevant term for pure gauge theory is the genus $0$ one, whether the higher genera free energies compute gravitational corrections. The topological strings compute an F-term of the gauge theory that can be written in function the ``glueball'' superfield $S:= \Tr \Psi_\alpha \Psi^\alpha$, where $\Psi^\alpha$ is the gluino field. If we define
\begin{eqnarray}
F:= \sum_{h=0}^\infty F_{0,h} S^h
\end{eqnarray}
then the F-term computed by the genus-0 topological string in the $\mathcal{N}=1$ theory can be written:
\begin{eqnarray}
\int \de^4 x \int \de^2 \theta N \frac{\partial F}{\partial S}
\end{eqnarray}
and one also has to include the term
\begin{eqnarray}
\int \de^4 x \int \de^2 \theta \tau S
\end{eqnarray}
i.e. the super Yang--Mill action in superfield notation, with the complexified gauge coupling
\begin{eqnarray}
\tau := \frac{4 \pi \ii}{g_{YM}^2} + \frac{\theta}{2 \pi}
\end{eqnarray}
and one defines the {\emph glueball superpotential}
\begin{eqnarray}
  W(S) := N \frac{\partial F}{\partial S} + \tau S
\end{eqnarray}
In the infra-red regime the glueball is conjectured to have a expectation value such that $W'(S)=0$, so that the vacuum structure is determined by the glueball superpotential.

\subsection{The closed side with fluxes}

To determine the contribution of topological strings, let us consider again Type IIB strings compactified on a Calabi--Yau manifold $X$. The prepotential term is given in $\mathcal{N}=2$ notation as
\begin{eqnarray}\label{eq:prepotential}
  \int \de^4 x \int \de^4 \theta F_0(X^I)
\end{eqnarray}
where $F_0$ is the genus $0$ B-model topological string free energy and the $X_I$ are the vector superfields, whose lowest components parametrize the complex structure of $X$. We now introduce $N^I$ units of the Ramond--Ramond $3$-form flux on the $I$-th A-cycle of $X$, where we have chosen a splitting of $H_3(X)$ into A-cycles and B-cycles with $X^I$ the periods of the A-cycles. In the language of $\mathcal{N}=2$ supergravity, the flux corresponds to the $\theta^2$ component of the superfield $X^I$. In this way, a vacuum expectation value absorbs two integrals in $\theta$ in (\ref{eq:prepotential}) and generate an F-term in the $\mathcal{N}=1$ language \cite{Vafa:2000wi} of the form
\begin{eqnarray}
\int \de^4 x \int \de^2 \theta N^I \frac{\partial F_0}{\partial X^I}
\end{eqnarray}
i.e. a superpotential for the moduli $X^I$.
If one consider also a flux $\tau_I$ on the I-th B-cycle, one obtains \cite{Gukov:1999ya,Taylor:1999ii}
\begin{eqnarray}
W(X^I) = N^I \frac{\partial F_0}{\partial X^I} + \tau_I X^I .
\end{eqnarray}

\chapter[String field theory and holomorphic Chern--Simons theory]%
{String field theory and\\ holomorphic Chern--Simons theory}%
\label{appendix-topologicalSFT}

In this appendix we briefly summarize the relation between open B-model and holomorphic Chern-Simons theory. In the first section we give a short derivation of this relation, following \cite{Marino:2004uf}. In the second section we discuss some properties of the holomorphic Chern-Simons functional and of its ``quantum theory''.

\section{The string field theory for the B-model}
The worldsheet $\sigma$--model description of open string theories is given by maps $\Sigma_{g,h} \to X$ from the open worldsheet, i.e. a Riemann surface $\Sigma_{g,h}$ of genus $g$ with $h$ boundaries, and the target space, a Riemannian manifold $X$. To define the theory one has also to specify boundary conditions for the fields. For the open B--model \cite{Witten:1992fb,Ooguri:1996ck}, the correct boundary conditions turn out to be Dirichlet along holomorphic cycles $S \subset X$ and Neumann in the remaining directions. It is also possible to add $N$ Chan--Paton factors and this corresponds to considering a $U(N)$ holomorphic bundle on the cycle $S$. These bondary conditions correspond to topological B-model in the presence of $N$ topological $D$ branes wrapping $S$. 
As in the closed case, the theory can be coupled to gravity. 

The target space description of this theory is given by the cubic string field theory \cite{Witten:1992fb}, in analogy to the bosonic string case \cite{Witten:1985cc}, see \cite{Taylor:2003gn} for a pedagogical introduction to the subject. In the bosonic cubic string field theory one considers the worldsheet of the string to be $\Sigma = \mathbb{R} \times I$, 
where $I=[0, \pi]$, with coordinates $0  \leq \sigma \leq \pi$ and $- \infty < \tau < \infty$ and a flat metric $\de s^2 = \de \tau^2 + \de \sigma^2$. One considers maps $x: I \to X$ where $X$ is the target space. The string field is a functional $\Psi[x(\sigma), \cdots]$ of open string configurations of ghost number $1$ and the $\cdots$ stands for the ghost fields (we will not indicate explicitly this dependence in the following). In the space of string functionals on defines two operations, the \emph{integration} and the \emph{star product}. The integration is defined by folding the string around its midpoint and gluing the two halves
\begin{eqnarray}
\int \Psi = \int \mathcal{D}x(\sigma) \prod_{0  \leq \sigma \leq \pi} \delta(x(\sigma) - x(\pi-\sigma)) \Psi[x(\sigma)]
\end{eqnarray}
and it has ghost number $-3$, i.e. the ghost number of the vacuum, corresponding to the fact that the open string theory on the disk one has to eliminate three zero modes. The associative and noncommutative star product $\star$ is defined as
\begin{eqnarray}
\int \Psi_1 \star \cdots \star \Psi_N := 	\qquad\qquad\qquad\qquad\qquad\qquad\qquad\qquad	\nonumber \\
                    \int \prod_{i=1}^{N}\mathcal{D} x_i(\sigma) \prod_{i=1}^{N} \prod_{0  \leq \sigma \leq \pi}
                    \delta \left( x_i(\sigma) - x_{i+1}(\pi - \sigma) \right) \Psi_i[x_i(\sigma)]
\end{eqnarray}                    
It glues the strings together by folding them around their midpoints and gluing the first half of one with the second half of the following and it does not change the ghost number. The cubic string field theory is defined as
\begin{eqnarray}
S:=\frac{1}{g_s} \int \left( \frac{1}{2} \Psi \star Q \Psi + \frac{1}{3} \Psi \star \Psi \star \Psi \right)
\end{eqnarray}
where $g_s$ is the string coupling constant. Since the integrand has ghost number $3$ and the integration has ghost number $-3$, the action has ghost number $0$. Adding Chan--Paton factors the string field become a $U(N)$ matrix of string fields and the integration includes a trace.

Topological strings have a structure analogous to the structure of bosonic strings,
namely a nilpotent BRST operator and an energy-momentum tensor that is BRST exact. For this reason 
one can use the same form for the string field action.
In this case one can prove that the higher modes decouple and only the zero modes contribute 
to the dynamics, see \cite{Witten:1992fb} and also the lectures \cite{Marino:2004uf}. For the B-model on a Calabi-Yau manifold, the string field turns out to be $A_{\bar I}$, a $(0,1)$-form taking values in the endomorphisms of a vector bundle $E$, i.e. the $(0,1)$ part of a gauge connection on $E$. 
The string field action is 
\begin{eqnarray}
S(\mathsf{A}) = \frac{1}{g_s} \int_X 
\Omega \wedge \Tr \left( \frac{1}{2}  \mathsf{A} \wedge \bar\partial \mathsf{A} + 
                         \frac{1}{3} \mathsf{A} \wedge \mathsf{A} \wedge \mathsf{A} \right) ,
\end{eqnarray}
the \emph{holomorphic Chern--Simons action}.

\section{The holomorphic Chern--Simons theory}\label{appendix-hCS}

Let us take a step back and let us take a closer look to the theory described by the holomorphic Chern--Simons functional, see \cite{Donaldson:1996kp,Thomas,Thomas:2000} and \cite{Nekrasov:2004vv}.
First notice that its minima are the connections satisfying 
\begin{eqnarray}
F^{(0,2)} = 0.
\end{eqnarray}
Having said that, two remarks are in order.%
The first one is that we are interested in a fuctional on $\mathcal{A}^{(0,1)}$ which is only ``almost'' gauge invariant,
up to ``large'' gauge transformations, in the following sense.
In general, given a smooth vector bundle $E \to X$, we may consider
the space of connections $\mathcal{A}$. On this space there is an action of the group of gauge transformations
$\mathcal{G}$. For $A \in \mathcal{A}$ and $g \in \mathcal{G}$ 
\begin{eqnarray}
A^{g} = g^{-1} A g + g^{-1} \de g .
\end{eqnarray}
When $E$ is a holomorphic vector bundle on a complex manifold $X$, we consider
the action of the complexified group of gauge transformations $\mathcal{G}_\mathbb{C}$. 
Given the complex structure on $X$ we decompose the connection $A$ as $A = A^{(1,0)} + A^{(0,1)}$
and for $g \in \mathcal{G}_\mathbb{C}$ we have the action (see \cite{DonaldsonKronheimer})
\begin{eqnarray}\label{eq:complexgauge}
\left\{\begin{array}{rcl}
   A^{(0,1)} &\to&   g^{-1} A^{(0,1)} g + g^{-1} \bar\partial g \\ 
   A^{(1,0)} &\to&   g^{-1*} A^{(1,0)} g^{*} + g^{-1*} \partial g^{*} 
\end{array}\right.
\end{eqnarray}
One can note that when $g=g^{*}$, one gets back the standard action of $\mathcal{G}$. 
Under the action of the complexified gauge group $\mathcal{G}_\mathbb{C}$, the holomorphic
Chern--Simons theory 
%
transforms as
\begin{eqnarray}
  S(\mathsf{A}^{g}) = S(\mathsf{A}) - \frac{1}{3} \int_X \Omega \wedge \left( g^{-1} \bar\partial g \right)^{3}
\end{eqnarray}
up to a possible numerical factor. The second term is a deformation invariant term, 
but not an integer for compact nonsingular Calabi--Yau manifolds, as discussed for example in \cite{Thomas}.

The second remark is connected to the fact that holomorphic Chern--Simons is actually a chiral (or holomorphic) theory and one should provide the ``slice'' in the space of fields over which one is integrating the ``measure''.

\chapter{Some definitions in geometry}\label{appendix-geometry}

In this appendix we will briefly recall some basic definitions. 
The classical references are \cite{Griffiths-Harris} 
for the analytic viewpoint and \cite{Hartshorne} for the algebraic one.
In particular, in the section \ref{appendix-families} we summarize 
the properties of families of complex submanifolds of a complex manifold, 
following \cite{namba}, see \cite{Kodaira} for the general theory 
of complex structure deformations and also the lecture notes \cite{Manetti:lectures}.

\section{Complex manifolds and vector bundles}
\begin{defin}
A \emph{complex manifold} $X$ is a differentiable manifold 
admitting an open cover $\{ U_\alpha \}$ and coordinate maps
$\phi_\alpha : U_\alpha \to \mathbb{C}^n$ such that 
$\phi_\alpha \circ \phi^{-1}_\alpha$ is holomorphic on $\phi_\beta
( U_\alpha \cap U_\beta ) \subset \mathbb{C}^n $ for all $\alpha, \beta$.
\end{defin}

Complex (connected) manifolds of complex dimension $1$ are called \emph{Riemann surfaces}; compact Riemann surfaces 
are all algebraic, i.e. they are the set of zeroes of systems of homogeneous polynomials. 

Given a complex manifold $X$, a \emph{complex submanifold} of $X$ is a pair
$(Y, \imath)$, where $Y$ is a complex manifold, and $\imath : Y \to X$ is an injective holomorphic map
whose jacobian matrix has rank equal to the dimension of $Y$ at any point of $Y$.

By the identification $\mathbb{C}^n \simeq \mathbb{R}^{2n}$, 
and since a biholomorphic map is a $C^\infty$ diffeomorphism, an $n$-dimensional complex manifold $X$
has an underlying structure of 2n-dimensional real manifold. 
Let $TX$ be the smooth tangent bundle. If $(z_1, \dots , z_n)$ is
a set of local complex coordinates around a point $x \in X$, then the complexified tangent
space $T_x X \otimes_\mathbb{R} \mathbb{C}$ admits the basis
\begin{eqnarray}
   \left( \left(\frac{\partial}{\partial z^1} \right)_x , \dots, \left(\frac{\partial}{\partial z^n} \right)_x ,
          \left(\frac{\partial}{\partial \bar{z}^1} \right)_x , \dots, 
                                           \left(\frac{\partial}{\partial \bar{z}^n} \right)_x \right) .
\end{eqnarray}
This yields a decomposition
\begin{eqnarray}
TX \otimes \mathbb{C} = T'X \oplus T''X
\end{eqnarray}
which is intrinsic because $X$ has a complex structure, so that the transition functions
are holomorphic and do not mix the vectors $\frac{\partial}{\partial z^i}$ with the 
$\frac{\partial}{\partial \bar{z}^i}$. As a consequence one has a decomposition
\begin{eqnarray}
  \Lambda^i T^* X \otimes \mathbb{C} = \bigoplus_{p+q = i} \Omega^{p,q} X \quad 
  \mathrm{where} \quad
  \Omega^{p,q} X = \Lambda^p(T'X)^* \otimes \Lambda^p(T''X)^*.
\end{eqnarray}
The elements in $\Omega^{p,q} X$ are called \emph{differential forms of type $(p, q)$}, and can locally be
written as
\begin{eqnarray}
   \eta = \eta_{i_1 \dots i_p, j_1 \dots j_q} (z, \bar{z}) 
            \de z^{i_1} \wedge \cdots \wedge \de z^{i_p} \wedge 
            \de \bar{z}^{j_1} \wedge \cdots \wedge \de \bar{z}^{j_1}
\end{eqnarray}
The highest exterior power of the holomorphic cotangent bundle 
\begin{eqnarray}
  K_X := \Lambda^n T'^{*}_X
\end{eqnarray}
is called the \emph{canonical bundle} of the $n$-dimensional complex manifold $X$. 
Holomorphic sections of $K_X$ are holomorphic $n$-forms.

\begin{defin}
An \emph{almost complex manifold} is a differential manifold $M$ with an endomorphism 
of the tangent bundle $J : TM \to TM$ such that $J^2 = -1$.
\end{defin}

\begin{proposition}
A complex manifold admits a natural almost complex structure.
\end{proposition}

Let $X$ be a differential manifold.
\begin{defin}
A $C^{\infty}$ \emph{complex vector bundle} on $X$ is given by a family $\{E_x\}_{x \in X}$
of complex vector spaces parametrized by $X$ together with a $C^{\infty}$ manifold
structure on $E = \bigcup_{x \in X} E_x$ such that the projection $\pi: E \to X$ 
mapping $E_x$ to $x$ is $C^{\infty}$ and for every $x_0 \in X$ there exists an open set
$U \subset X$ containig $x_0$ and
\begin{eqnarray}
\phi_U : \pi^{-1}(U) \to U \times \mathbb{C}^k \nonumber
\end{eqnarray}
taking $E_x$ (linearly and) isomorphically onto $\{x\} \times \mathbb{C}^k$ for any $x \in U$.
$\phi_U$ is called a \emph{trivialization} of $E$ over $U$.
\end{defin}

Let $X$ be a complex manifold.
\begin{defin}
A \emph{holomorphic vector bundle} $\pi : E \to X$ is a complex vector bundle
together with the structure of a complex manifold on $E$ such that for any
$x \in X$ there exists $U$ containing $x$ in $M$ and a trivialization
\begin{eqnarray}
   \phi_U: E_U \to U \times \mathbb{C}^k \nonumber
\end{eqnarray}
that is a biholomorphic map of complex manifolds.
\end{defin}

\section{Submanifolds and their families}%
\label{appendix-families}

Let $X$ be an $(r+d)$--dimensional (connected) complex manifold and $S \subset X$ a $d$--dimensional (connected) compact complex submanifold. We may assume that $S$ is covered by a finite number of open subsets $\{W_\alpha\}_{\alpha \in I}$ of $X$ each of which has a local coordinate system
\begin{eqnarray}
(z_\alpha, w_\alpha) = (z_\alpha^1, \dots, z_\alpha^d, w_\alpha^1, \dots w_\alpha^r) \nonumber
\end{eqnarray}
such that $S$ is defined in $W_\alpha$ by $w_\alpha = 0$. We define $U_\alpha := W_\alpha \cap S$. On $W_{\alpha \beta}:= W_{\alpha} \cap W_{\beta}$ we have the coordinate transformations
\begin{eqnarray}
\left\{\begin{array}{rcl}
z_\alpha &=& f_{\alpha \beta} (z_\beta, w_\beta) \\
w_\alpha &=& g_{\alpha \beta} (z_\beta, w_\beta)
\end{array}\right.
\end{eqnarray}
where $f_{\alpha \beta} = (f^1_{\alpha \beta}, \dots, f^d_{\alpha \beta})$ and $g_{\alpha \beta} = (g^1_{\alpha \beta}, \dots, g^r_{\alpha \beta})$ are  vector-valued holomorphic functions of $(z_\beta, w_\beta) \in W_{\alpha \beta}$.

Let us define the matrix--valued holomorphic functions
\begin{eqnarray}
N_{\alpha \beta} := \left( \frac{\partial g_{\alpha \beta}}{\partial w_\beta} \right)_{(z_\beta, 0)} 
\qquad z_\beta \in U_{\alpha \beta} \nonumber
\end{eqnarray}
We then have the following compatibility conditions
\begin{eqnarray}
N_{\alpha \beta}(z_\beta) N_{\beta \gamma}(z_\gamma) = N_{\alpha \gamma}(z_\gamma)
\qquad \mathrm{for} \ z_\gamma \in U_{\alpha \beta \gamma} \ \mathrm{and} \ z_\alpha:=f_{\alpha \beta}(z_\beta, 0)
\end{eqnarray}
Thus $\{N_{\alpha \beta}\}_{\alpha,\beta \in I}$ defines a holomorphic vector bundle $N \to S$, called the \emph{normal bundle} of $S$ in $X$, i.e. the quotient bundle
\begin{eqnarray}
0 \to TS \to TX|_S \to N_{S|X} \to 0
\end{eqnarray}

Let now $S'\subset X$ be another compact complex submanifold of $X$ covered by $\{W_\alpha\}_{\alpha \in I}$ and assume that $S'$ is defined in $W_\alpha$ by the equations
\begin{eqnarray}
w_\alpha = \phi_\alpha(z_\alpha) \nonumber
\end{eqnarray}
where $\phi_\alpha$ is a vector valued holomorphic function of $z_\alpha \in U_\alpha$. These $\phi_\alpha$ have to satisfy the compatibility condition
\begin{eqnarray}
g_{\alpha \beta}(\phi_\beta(z_\beta), z_\beta) = \phi_\alpha(f_{\alpha \beta}(\phi_\beta(z_\beta)), z_\beta) \qquad \mathrm{for} \ (\phi_\beta(z_\beta), z_\beta) \in W_{\alpha \beta}
\nonumber
\end{eqnarray}
We want to consider families of such $S'$. Let us first introduce the notion of complex analytic space, a generalization of the concept of a complex analytic manifold.
An \emph{analytic space} over $\mathbb{C}$ is a ringed space that is locally isomorphic to a ringed space 
$(X, \mathcal{O})$, where $X$ is defined in a domain $U$ of $\mathbb{C}^n$ by equations $f_1 = \dots = f_p = 0$, $f_i$ are analytic functions on $U$, and $\mathcal{O}$ is obtained by restricting the sheaf $\mathcal{O}_U / I$ on $X$; here $\mathcal{O}_U$ is the sheaf of germs of analytic functions in $U$, while $I$ is the subsheaf of ideals generated by  $f_1, \dots, f_p$. 

\begin{defin}
Let $M$ and $B$ be analytic spaces (i.e. reduced, connected, Hausdorff complex analytic spaces) and $\phi: M \to B$ a proper surjective holomorphic map. The triple $(M, \pi, B)$ is called a \emph{family of compact complex manifolds} if there exist $\{M_\alpha\}$ open covering of $M$, $\Omega_\alpha \subset \mathbb{C}^n$ open subsets and holomorphic isomorphisms
\begin{eqnarray}
\eta_\alpha: M_\alpha \to \Omega_\alpha \times B_\alpha \nonumber
\end{eqnarray}
where $B_\alpha:=\pi(M_\alpha)$ is open in $B$, such that the diagram
$$\xymatrix{
M_\alpha \ar[dr]_\pi \ar[rr]^{\eta_\alpha} &   & {\Omega_\alpha \times B_\alpha}  \ar[dl]^{\mathrm{proj}}\\
                 & B_\alpha &}$$
commutes for each $\alpha$. $B$ is called the \emph{parameter space} of the family $(M, \pi, B)$.
\end{defin}

\begin{defin}
Let $X$ be a complex manifold. A family of compact complex manifolds $(M,\pi,B)$ is called a \emph{family of compact complex submanifolds} of $X$ if 
\begin{itemize}
\item $M$ is an analytic subvariety of $X \times B$
\item $\pi$ is the restriction to $M$ of $\mathrm{proj}:X\times B \to B$.
\end{itemize}
\end{defin}

For each $b \in B$ of a family $(M,\pi,B)$ of submanifolds of $X$, the fibre $\pi^{-1}(b)$ can be written as $\pi^{-1}(b) = S_b \times b$ where $S_b$ is a compact complex submanifold of $X$, so we identify $\pi^{-1}(b)$ with $S_b$ and write the family as $\{S_b\}_{b \in B}$.

\begin{defin}
A family $\{S_b\}_{b \in B}$ of compact complex submanifolds of $X$ is said to be \emph{maximal} at $b_0 \in B$ if for each $\{S_c\}_{c \in C}$ family of compact complex submanifolds of $X$ with a point $c_0$ such that $S_{c_0} = S_{b_0}$ there exists $U$ neighbourhood of $c_0$ in $C$ and a holomorphic map $f: U \to B$ such that $f(c_0)=b_0$ and $S_{f(c)}=S_c$ for each $c \in U$. $\{S_b\}_{b \in B}$ is said to be a \emph{maximal family} if it is maximal at every point of the parameter space.
\end{defin}

We then have the following results, \cite{namba}.

\begin{thm}
There exists a maximal family $\{S_b\}_{b \in B}$ of compact complex submanifolds of $X$ such that $S_o = S$ for a point $o \in B$ where the parameter space $B$ is an analytic space.
\end{thm}
Fixing $X$, one can patch these families together and obtain the following theorem.
\begin{thm}
Let $X$ be a complex manifold. Then the set of all compact complex submanifolds of $X$ forms a (not necessarily connected) analytic space $B(X)$ in a natural way.
\end{thm}


\section{Connections and hermitian metrics}

\begin{defin}
Let $X$ be a differential manifold and $E \to X$ a complex vector bundle. A \emph{connection} is a map
\begin{eqnarray}
\nabla : \Omega^0(E) \to \Omega^1(E) \nonumber
\end{eqnarray}
satisfying the Leibnitz rule 
\begin{eqnarray}
  \nabla (f s) = f \nabla (s) + \de f \otimes s \nonumber
\end{eqnarray}
for every section $s$ of $E$ and every function $f$ on $X$ or on any open subset.
\end{defin}

The Leibniz rule also shows that $\nabla$ is $\mathbb{C}$-linear. 
The connection $\nabla$ can be made to act on all
$\Omega^k (E)$, thus getting a morphism
\begin{eqnarray}
\nabla : \Omega^k(E) \to \Omega^{k+1} (E) \nonumber
\end{eqnarray}
by letting 
\begin{eqnarray}
  \nabla (\omega \otimes s) = \de \omega \otimes s +
                              (-1)^k \omega \otimes \nabla (s) .\nonumber
\end{eqnarray}

If $\{U_\alpha\}$ is a cover of $X$ over which $E$ trivializes, we may choose on any $U_\alpha$ a set
$\{s_\alpha \}$ of basis sections of $E_{U_\alpha}$ (notice that this is a set of $r$ sections, with $r = \rk E$).
Over these bases the connection $\nabla$ is locally represented by matrix-valued differential 1-forms $A_\alpha$:
\begin{eqnarray}
  \nabla (s_\alpha) = A_\alpha \otimes s_\alpha .\nonumber
\end{eqnarray}
Every $A_\alpha$ is as an $r \times r$ matrix of 1-forms. The $A_\alpha$'s are called \emph{connection 1-forms}.

If $g_{\alpha \beta}$ denotes the transition functions of $E$ with respect
to the chosen local basis sections (i.e., $s_\alpha = g_{\alpha \beta} s_\beta$), the transformation formula for the
connection 1-forms is
\begin{eqnarray}
  A_\alpha = g_{\alpha \beta} A_{\beta} g_{\alpha \beta}^{-1} + \de g_{\alpha \beta} g_{\alpha \beta}^{-1}.
\end{eqnarray}
The connection is not a tensorial morphism, but rather satifies a Leibniz rule; as a
consequence, the transformation properties of the connection 1-forms are inhomogeneous
and contain an affine term.

The square of the connection
\begin{eqnarray}
  \nabla^2 : \Omega^k(E) \to \Omega^{k+2}(E)
\end{eqnarray}
is $f$-linear, i.e. it satisfies the property $\nabla^2(fs)=f\nabla^2(s)$
for every function $f$ on $X$. In other terms, $\nabla^2$ is an endomorphism of the bundle $E$
with coefficients in 2-forms, namely, a global section of the bundle $\Omega^2 \otimes \End(E)$. 
It is called the \emph{curvature} of the connection $\nabla$ and we shall denote it by $F$. 
On local basis sections $s_\alpha$ it is represented by the \emph{curvature 2-forms} $F_\alpha$ defined by
\begin{eqnarray}
  F(s_\alpha) = F_\alpha \otimes s_\alpha .
\end{eqnarray}

The curvature 2-forms may be expressed in terms of the
connection 1-forms by the equation (Cartan's structure equation)
\begin{eqnarray}
  F_\alpha = \de A_\alpha - A_\alpha \wedge A_\alpha
\end{eqnarray}
and the transformation formula for the curvature 2-forms is
\begin{eqnarray}
  F_\alpha = g_{\alpha \beta} F_\beta g_{\alpha \beta}^{-1} .
\end{eqnarray}
Due to the tensorial nature of the curvature morphism, the curvature 2-forms obey a
homogeneous transformation rule, without affine term.

\paragraph{}
An Hermitian metric $h$ of a complex vector bundle $E$ is a global section of $E \otimes \bar{E^*}$ which when
restricted to the fibres yields a Hermitian form on them.
On a local basis of sections of $\{s_\alpha\}$ of $E$, $h$ is represented by matrices $h_\alpha$ of functions on
$U_\alpha$ which are Hermitian and positive definite at any point of $U_\alpha$.
The local basis is said to be \emph{unitary} if the corresponding matrix h is
the identity matrix.

A pair $(E, h)$ formed by a holomorphic vector bundle with a hermitian metric is
called a \emph{hermitian bundle}. A connection $\nabla$ on $E$ is said to be \emph{metric} with respect to $h$
if for every pair $s, t$ of sections of E one has
\begin{eqnarray}
\de h(s, t) = h(\nabla s, t) + h(s,\nabla t) .
\end{eqnarray}
In terms of connection forms and matrices representing $h$ this condition reads
\begin{eqnarray}
 \de h_\alpha = A_\alpha^{t} h_\alpha + h_\alpha \bar{A}_\alpha
\end{eqnarray}
where ${}^t$ denotes transposition and the bar denotes complex conjugation. This equation implies that on a
unitary frame, the connection forms are skew-hermitian matrices.

\begin{proposition}
Given a hermitian bundle $(E, h)$, there is a unique connection $\nabla$
on $E$ which is metric with respect to $h$ and is compatible with the holomorphic structure
of $E$.
In a holomorphic local basis of sections, the connection forms are 
\begin{eqnarray}
  A_\alpha^{t} = \partial h_\alpha h_\alpha^{-1} .
\end{eqnarray}
\end{proposition}

The connection of the previous proposition is called \emph{Chern connection}.

\section{K\"ahler manifolds and Calabi--Yau manifolds}

\begin{defin}
Let $X$ be a complex manifold with complex structure $J$ and $h$ a hermitian metric on $X$. The form
\begin{eqnarray}
  \omega (X, Y) := \left< J X, Y \right> = - \left< X, JY \right> \qquad \mathrm{for} \qquad X, Y \in \Gamma(TX)
\end{eqnarray}
is called \emph{hermitian form} (or \emph{fundamental form}).
\end{defin}

\begin{defin}
A complex manifold with hermitian metric $h$ and hermitian form $\omega$ is called a \emph{K\"ahler manifold} if 
$ \de \omega = 0$.
\end{defin}

\begin{defin}
A \emph{Calabi--Yau manifold} is a K\"ahler manifold with trivial canonical bundle.
\end{defin}

Actually there are many nonequivalent definitions of a Calabi--Yau manifold, we just took a reasonable one. 
Notice that one often relaxes the K\"ahler condition, as we did. Notice also that this is slightly cheating, since of course one is interested also in (not too much) singular spaces.

\section{Direct and inverse image sheaves}\label{appendix-sheaves}

In this section we list a few definitions and results from \cite{Hartshorne}. 
We shall assume, among the others, the definitions of sheaf, scheme and 
\v{C}ech cohomology groups.

\begin{defin}
Let $f: X \to Y$ a continuous map of topological spaces and $\mathcal{F}$ a sheaf on $X$. 
The \emph{direct image sheaf} $f_* \mathcal{F}$ on $Y$ is defined by
\begin{eqnarray}
  ( f_* \mathcal{F} )(V):= \mathcal{F} (f^{-1}(V)) .\nonumber
\end{eqnarray}
Let now $\mathcal{G}$ be a sheaf on $Y$. The \emph{inverse image sheaf} on $X$ 
is the sheaf associated to the presheaf 
\begin{eqnarray}
  U \to \lim_{V \supseteq f(U)} \mathcal{G} (V) \nonumber
\end{eqnarray}
where $U \subseteq X$ is an open set in $X$ and the limit is taken over all 
open sets $V \subseteq Y$ containing $f(U)$.
\end{defin}

Let now $f: (X, \mathcal{O}_X) \to (Y, \mathcal{O}_Y)$ be a morphism of ringed spaces.
If $\mathcal{F}$ is a sheaf of $\mathcal{O}_X$-modules, then $f_* \mathcal{F}$ 
is a sheaf of $f_*\mathcal{O}_X$-modules.
The morphism $f^\sharp : \mathcal{O}_Y \to f_*\mathcal{O}_X$ of sheaves of rings on $Y$
gives $f_* \mathcal{F}$ a natural structure of $\mathcal{O}_Y$-module.

\begin{defin}
  The \emph{direct image sheaf} $f_* \mathcal{F}$ of $\mathcal{F}$ by the morphism $f$ is the 
  sheaf of $\mathcal{O}_Y$-modules constructed above.
\end{defin}

If $\mathcal{G}$ is a sheaf of $\mathcal{O}_Y$-modules, then $f^{-1} \mathcal{G}$ 
is a sheaf of $f^{-1}\mathcal{O}_Y$-modules. Notice that we have a morphism 
$f^{-1} \mathcal{O}_Y \to \mathcal{O}_X$ of sheaves of rings on $X$.

\begin{defin}
  The \emph{inverse image sheaf} of $\mathcal{G}$ by the morphism $f$ is the 
  sheaf of $\mathcal{O}_X$-modules
  \begin{eqnarray}
    f^* \mathcal{G} := f^{-1} \mathcal{G} \otimes_{f^{-1} \mathcal{O}_Y} \mathcal{O}_X . \nonumber
  \end{eqnarray}
\end{defin}

\begin{proposition}
  Let $f: X \to Y$ be a morphism of schemes. Then
  \begin{enumerate}
    \item if $\mathcal{G}$ is a quasi-coherent sheaf of $\mathcal{O}_Y$-modules, 
          then $f^* \mathcal{G}$ is a quasi-coherent sheaf of $\mathcal{O}_X$-modules;
    \item if $X$ and $Y$ are noetherian and $\mathcal{G}$ is coherent,
          then $f^* \mathcal{G}$ is coherent;
    \item if $X$ is noetherian or $f$ is quasi-compact and separated and 
          and $\mathcal{F}$ is a quasi-coherent sheaf of $\mathcal{O}_X$-modules,
          then $f_* \mathcal{F}$ is a quasi-coherent sheaf of $\mathcal{O}_Y$-modules.
  \end{enumerate}
\end{proposition}

\begin{proposition}[Projection formula]\label{prop:projection}
Let $f: (X, \mathcal{O}_X) \to (Y, \mathcal{O}_Y)$ be a morphism of ringed spaces,
$\mathcal{F}$ a sheaf of $\mathcal{O}_X$-modules and
$E$ a locally free sheaf of $\mathcal{O}_Y$-modules of finite rank.
Then there exists a natural isomorphism
\begin{eqnarray}
  f_* \left( \mathcal{F} \otimes_{\mathcal{O}_X} f^* E \right) 
      \simeq f_* ( \mathcal{F} ) \otimes_{\mathcal{O}_Y} E . \nonumber
\end{eqnarray}
\end{proposition}
In particular, taking $\mathcal{F} \simeq \mathcal{O}_X$ in the previous proposition, we obtain
\begin{eqnarray}
  f_* f^* E \simeq f_* ( \mathcal{O}_X ) \otimes_{\mathcal{O}_Y} E . 
\end{eqnarray}

\begin{defin}
  A morphism of schemes $f: X \to Y$ is an \emph{affine morphism} if
  there exists an open affine cover $\mathcal{V} = \{ V_\alpha \}$ of $Y$ 
  such that $f^{-1}(V_\alpha)$ is affine for any $\alpha$.
\end{defin}

Any affine morphism is quasi-compact and separated.

\begin{proposition}\label{prop:affine}
  Let $f: X \to Y$ be an affine morphism of noetherian separated schemes and
  $\mathcal{F}$ a quasi-coherent sheaf on $X$. Then there are natural isomorphisms
  \begin{eqnarray}
    H^i ( X, \mathcal{F} ) \simeq H^i (Y, f_* \mathcal{F}), \qquad i \geq 0 . \nonumber
  \end{eqnarray}
\end{proposition}


\end{document}